\documentclass[12pt]{article}

\usepackage[english]{babel}

\usepackage{setspace}
\usepackage{csquotes}

\usepackage{adjustbox}

\usepackage[letterpaper,top=2cm,bottom=2cm,left=2.5cm,right=2.5cm,marginparwidth=1.75cm]{geometry}

\usepackage{amsmath}
\usepackage{graphicx}
\usepackage[colorlinks=true, allcolors=blue]{hyperref}

\usepackage{url,enumitem}
\usepackage{amssymb}
\usepackage{subcaption}
\usepackage{pdflscape}
\usepackage{changepage}
\usepackage{bm}
\usepackage{booktabs}
\usepackage{tabularx}

\usepackage[format=hang,labelfont=bf]{caption}
\usepackage[figuresright]{rotating}
\usepackage{multirow}
\usepackage{hhline}
\usepackage{enumerate}
\usepackage{verbatim}
\usepackage{colortbl}
\usepackage{tabularx}
\usepackage{rotating}
\usepackage{threeparttable}
\usepackage{authblk}
\usepackage{footnote}
\usepackage{hyperref}
\usepackage[style=numeric-comp, sorting=none]{biblatex}
\title{Interrater agreement statistics under the two-rater dichotomous-response case with correlated decisions}

\author[1]{Zizhong Tian, MS}
\author[1]{Vernon M. Chinchilli, PhD}
\author[1,2]{Chan Shen, PhD}
\author[1]{Shouhao Zhou, PhD\thanks{Corresponding author. Email: szhou1@pennstatehealth.psu.edu}}
\affil[1]{Department of Public Health Sciences, Pennsylvania State University, Hershey, PA, USA}
\affil[2]{Department of Surgery, Pennsylvania State University, Hershey, PA, USA}

\date{}
\begin{document}
\maketitle

%\begin{spacing}{1.5}
\subsection*{Summary}
Measurement of the interrater agreement (IRA) is critical in various disciplines. To correct for potential confounding chance agreement in IRA, Cohen's $\kappa$ and many other methods have been proposed. However, owing to the varied strategies and assumptions across these methods, there is a lack of practical guidelines on how these methods should be preferred even for the common two-rater dichotomous rating. To fill the gaps in the literature, we systematically review nine IRA methods and propose a generalized framework that can simulate the correlated decision processes behind the two raters to compare those reviewed methods under comprehensive practical scenarios. Based on the new framework, an estimand of ``true'' chance-corrected IRA is defined by accounting for the ``probabilistic certainty'' and serves as the comparison benchmark. We carry out extensive simulations to evaluate the performance of the reviewed IRA measures, and an agglomerative hierarchical clustering analysis is conducted to assess the interrelationships among the included methods and the benchmark metric. Recommendations for selecting appropriate IRA statistics in different practical conditions are provided and the needs for further advancements in IRA estimation methodologies are emphasized.

\textit{Keywords}: interrater agreement; dichotomous response; Monte Carlo simulation; correlated binary variables.
%\end{spacing}

\section{Introduction}

Making disease diagnosis or other ratings for a group of patients or experimental subjects based on certain judgement criteria is a common practice in psychology, radiology, education, and other relevant research areas \parencite{kazdin2021research,m2020Interobserver}. Given that most rating tasks, especially those involving human-based judgments, may not perfectly reflect the unknown truth behind the rating subjects, and that decision-making can vary due to rater and subject characteristics or other random factors, the use of multiple raters is increasingly employed to enhance research quality. Without a gold standard, measuring the extent of agreement among different raters' judgments can help illuminate the reliability and usability of the rating results. Therefore, it is vital to develop concise yet informative measures to quantify the interrater agreement (IRA) for studies with multiple raters. To date, numerous IRA statistics have been developed to fulfill the needs in various scientific disciplines \parencite{gisev2013interrater}. These statistics not only enable researchers to assess the between-rater reliability of a single rating method but also provide a valuable tool for evaluating the exchangeability of two judgement criteria, for instance, a newly developed rating method versus a traditional decision scheme \parencite{Choudhary2005}. The application of IRA statistics offers valuable insights into the validity and quality of the judgement criteria in use.

Cohen's $\kappa$ \parencite{cohen1960coefficient} is the most popular IRA statistic in psychology and social science studies and its calculation has been built into most analytical software. Despite its wide use in different fields, $\kappa$ has received many criticisms for its abnormal performances when the prevalence of underlying outcome is extreme or when there is a large discrepancy between the two raters' marginal distributions of ratings \parencite{feinstein1990high, cicchetti1990high, vach2005dependence, zhao2013assumptions}. This challenges $\kappa$ as a regular tool, since for many rating cases, it is unable to estimate the prevalence or the marginal distribution of rating a priori. Though not known well by applied researchers, some alternative choices for $\kappa$ were either identified in relevant statistical or mathematical literature or developed in later work to overcome the shortcomings in $\kappa$. For instance, Yule's $Y$ \parencite{yule1912methods} was found to provide better protection against extremely unbalanced marginal totals than $\kappa$ \parencite{spitznagel1985proposed}, and Gwet's $\text{AC}_1$ \parencite{gwet2008computing, gwet2014handbook} is a relatively new IRA method designed to address the paradoxes faced by $\kappa$ and similar methods \parencite{bexkens2018kappa, zhao2013assumptions}. Additionally, constrained by limited human resource and training costs, assigning two qualified raters to conduct a series of binary-outcome rating tasks is the most common scenario in clinical practice \parencite{blackman1993estimating, bloch19892}. Motivated by the abundance of IRA statistics suitable for the two-rater dichotomous-response scenario, although many of them are rarely acknowledged in real practice, we conducted a systematic comparison among a family of IRA statistics that are popular in practice, aiming to help researchers identify the most appropriate choices under different practical conditions.

While Monte Carlo simulation is a scientific approach to assess various IRA statistics under the two-rater binary-response case, the number of simulation studies is limited. \textcite{gwet2008computing} proposed a simulation framework that can identify a ``true'' measure of IRA, which further served as a benchmark in his simulation-based method comparison. He concluded that the $\text{AC}_1$ estimator he developed outperformed Cohen's $\kappa$, Scott's $\pi$ \parencite{scott1955reliability}, and Holley and Guilford's $G$ \parencite{holley1964note} in terms of bias toward the proposed ``true'' measure. \textcite{grant2017evaluation} evaluated 5 IRA indices based on their correlations with $d$-prime, an indirect benchmark reflecting raters' abilities to distinguish the true presence or absence of the outcome. They proposed a data-generating mechanism that produced each rater's responses based on the true outcomes unknown to the raters as well as rater characteristics such as expertise, bias in voting distribution, and attention level, and they endorsed Cohen's $\kappa$ and the Phi coefficient, which is the Pearson's correlation coefficient estimated from two binary random variables \parencite{yule1912methods, marshall1985family}. \textcite{xu2014interrater} used a Monte Carlo simulation to compare 6 IRA statistics under different settings of raters' base rates, rater biases, and sample sizes. They ranked the candidate IRA methods based on three qualitative ``optimum'' criteria, and they concluded that Holley and Guilford's $G$ was the best choice, followed by Gwet's $\text{AC}_1$. Feng did two simulation studies without proposing any benchmark for quantitative comparison \parencite{feng2013underlying, feng2013factors}. Instead, he focused on exploring the impacts of rater or rating subject-related factors (e.g., marginal distributions, sensitivity and specificity, subject difficulty levels, etc.) on several widely-used IRA statistics using regression or correlation analyses. In \textcite{feng2013underlying}, the author simulated the rating responses on each subject by simulating the difficulty level of the subject first. Then, based on some pre-specified correct rates corresponding to different difficulty levels as well as the underlying true outcomes, the observed responses on each subject were generated. Particularly, all these existing simulation schemes regarded the raters as independent agents. In real multiple-rater rating tasks, raters working on the same group of patients tend to have naturally correlated rating behaviors, and the level of such correlation may vary based on rater training or special subject features. Failing to consider this underlying correlation in the simulation study may bias the evaluation of the IRA methods. To fill this gap, we propose a generalized simulation framework that can emulate the important underlying factors in real-world rater operations, and we conduct a simulation study to evaluate the performance of the IRA statistics included in our review. Further, this comparative simulation study offers the most comprehensive list of IRA methods under the two-rater dichotomous-response case by far, where $\kappa$ and a number of alternative IRA measures were evaluated under varied conditions, including the prevalence of the underlying outcome and various aspects of rater characteristics.

The remainder of this paper is organized as follows. Section \ref{sec: 2} reviews a list of IRA statistics belonging to a popular methodological framework, emphasizing the mathematical connections among their point estimates and presenting their variance estimation formulas. Section \ref{sec: 3} introduces a generalized framework for assessing IRA statistics, considering the potential rater correlations underneath the rating processes. We define and illustrate an estimand of IRA based on a probabilistic ``truth'' table, and the data-generating mechanism for the simulation study is outlined with mathematical reasoning. Section \ref{sec: 4} describes the simulation settings and presents the comparison results of the reviewed methods. Key findings and practical suggestions about the appropriate choices of IRA statistics under specific practical conditions are discussed with concluding remarks in Section \ref{sec: 5}.

\section{Chance-corrected interrater agreement statistics and their connections}\label{sec: 2}

Two-rater dichotomous-response data are often organized in a $2 \times 2$ count table (Table \ref{tb: 2x2tb}) with the sum of counts fixed to the total number of rating subjects $N$. To estimate IRA, a naive estimator is the observed proportion of agreement, $\hat{p}_a=\hat{p}_{11}+\hat{p}_{00}$. However, there has been a common critique that $\hat{p}_a$ does not account for the statistical randomness in the rating process, or the possibility of making agreement by chance \parencite{cohen1960coefficient, banerjee1999beyond, gwet2014handbook}. Subsequently, most of the IRA measures, like Cohen's $\kappa$, adopt the chance-correcting approach and share the same general formula to estimate
\begin{equation}\label{eq: cc-agree}
\text{chance-corrected agreement}=\frac{p_a-p_{e}}{1-p_{e}},
\end{equation}
where $p_a=p_{11}+p_{00}$ is the ``total agreement'' and $p_e$ is the quantity of ``chance agreement'' \parencite{fleiss1975measuring, landis1975review, blackman1993estimating}. The numerator and the denominator respectively stand for the ``agreement beyond chance'' and the ``maximum attainable amount of agreement beyond chance,'' and a good perspective to understand \eqref{eq: cc-agree} is to regard it as the adjusted proportion of agreement after chance agreement is excluded from consideration \parencite{cohen1960coefficient}. The upper bound of \eqref{eq: cc-agree} is $1$, suggesting perfect agreement. The lower bound carries the form of $-p_e/(1-p_e)$, with a value no greater than $0$. With $p_a$ fixed, \eqref{eq: cc-agree} is strictly decreasing with respect to $p_e$.

\begin{table}[htbp]%[!th]
\caption{$2 \times 2$ Contingency Table Based on Two-Rater Dichotomous-Response Rating}\label{tb: 2x2tb}
\begin{center}
%\resizebox{\textwidth}{!}{
\begin{tabular}{cccc} 
\toprule
 & & \multicolumn{2}{c}{Rater 2} \\ 
\cmidrule(lr){3-4}
 &   & $+$ & $-$ \\ 
\midrule
\multirow{2}{*}{Rater 1} & $+$ & $n_{11}$ ($\hat{p}_{11}=n_{11}/N$) & $n_{10}$ ($\hat{p}_{10}=n_{10}/N$)  \\
& $-$ & $n_{01}$ ($\hat{p}_{01}=n_{01}/N$) & $n_{00}$ ($\hat{p}_{00}=n_{00}/N$)  \\
\bottomrule
\end{tabular}%}
\end{center}
\small\textit{Note}. The subscripts respectively indicate the vote from Rater 1 and 2 (``1'' means ``$+$'' and ``0'' means ``$-$''). For example, $n_{10}$ denotes the count that Rater 1 votes ``$+$'' but Rater 2 votes ``$-$'', and $\hat{p}_{10}=n_{10}/N$ denotes the corresponding proportion in the observed data sample. It follows $N=n_{11}+n_{10}+n_{01}+n_{00}$ and $\hat{p}_{11}+\hat{p}_{10}+\hat{p}_{01}+\hat{p}_{00}=1$.
\end{table}

\subsection{Point estimation}

In statistical language, \textit{estimands} are the expressions consisting of the population true parameter, while \textit{estimators} are the quantities expressed based on the observed data sample, targeting to estimate some true parameter or estimand. For instance, behind the observed Table \ref{tb: 2x2tb}, the true probability that Rater 1 votes ``$+$'' is $p_{1.}=p_{11}+p_{10}$, which is an \textit{estimand}; on the other hand, the \textit{estimator}, denoted as $\hat{p}_{1.}$, can be constructed as $\hat{p}_{11}+\hat{p}_{10}=(n_{11}+n_{10})/N$. Different IRA statistics, even though having the same goal of describing the intangible agreement level, have different \textit{estimands} on their own, reflecting diverse perceptions of the concept of ``agreement'' and different assumptions imposed when designing those IRA measurement tools. However, in real applications, the researchers will have more interest on the \textit{estimators} of the IRA measures and their performance. Therefore, our review in Section \ref{sec: 2} and the comparative study later only discuss how the \textit{estimators} were constructed under different IRA methods and how these estimators perform under different scenarios.

For those IRA statistics taking the form of \eqref{eq: cc-agree}, all of them build up their expressions by plugging in the respective estimators for $p_a$ and $p_e$. All these chance-corrected IRA statistics use the observed proportion of agreement ($\hat{p}_a$) to estimate total agreement, but they make different assumptions when constructing the estimator for chance agreement ($\hat{p}_e$). 

\textcite{scott1955reliability} proposed the $\pi$ statistic under the assumption of rater homogeneity and rater independence, meaning that the two raters follow the same marginal distribution of operation and they vote independently. Assume both raters have probability $\omega$ to vote for ``$+$'' and probability $(1-\omega)$ for ``$-$'', then the chance-agreement part in Scott's $\pi$ can be estimated by
\begin{align*}
    \hat{p}_{e|\pi}
    &=\hat{\omega}^2+(1-\hat{\omega})^2\\
    &=\left(\frac{2n_{11}+n_{10}+n_{01}}{2N}\right)^2+\left(\frac{2n_{00}+n_{01}+n_{10}}{2N}\right)^2,
\end{align*}
where $\hat{\omega}=(2n_{11}+n_{10}+n_{01})/(2N)$ is the estimated average probability to vote for category ``$+$''. Under the two-rater dichotomous-response or the $2\times 2$ case, Scott's $\pi$ is equivalent to the maximum likelihood estimator of Bloch and Kraemer's intraclass kappa \parencite{bloch19892}, which is a measure derived from a population model and by analogy with the intraclass correlation coefficient. In addition, %if discussing in the perspective of the estimand design, then the 
$\pi$ is equivalent to the Phi coefficient ($\phi$) estimated from two \textit{exchangeable} binary random variables. In a general setting, $\phi$ estimated from two raters' binary responses is given as
\begin{align*}
    \phi=\frac{\hat{p}_{11}-\hat{p}_{1.}\hat{p}_{.1}}{\sqrt{\hat{p}_{1.}(1-\hat{p}_{1.})\hat{p}_{.1}(1-\hat{p}_{.1})}}.
\end{align*}
Then, when the two raters have the same estimated marginal probabilities (i.e., $\hat{p}_{1.}=\hat{p}_{.1}=\hat{\omega}$), or equivalently, $n_{10}=n_{01}$, the following equality between the $\pi$ estimate and $\phi$ estimate can be shown,
\begin{align*}
    \phi=\frac{\hat{p}_{11}-\hat{\omega}^2}{\sqrt{\hat{\omega}^2(1-\hat{\omega})^2}}=\frac{\hat{p}_{11}+\hat{p}_{00}-\hat{\omega}^2-(1-\hat{\omega})^2}{1-\hat{\omega}^2-(1-\hat{\omega})^2}=\pi.
\end{align*}

Krippendorf's $\alpha$ \parencite{krippendorff1970bivariate, hayes2007answering} defined the chance agreement similarly based on the assumption that the raters were exchangeable and independently operated following identical marginal distributions. %but it additionally included a sense of ``without replacement'' 
When estimating the chance agreement (the joint probability of agreement conditional on independence), instead of using the probability estimates from the $2\times 2$ table and direct multiplication to get the joint agreement probability, $\alpha$ estimates the probability of agreeing on each category (``$+$'' or ``$-$'') by treating the two raters as a whole and incorporating a sense of drawing without replacement. The corresponding chance agreement is explicitly estimated by
\begin{equation*}
    \hat{p}_{e|\alpha}=\left(\frac{2n_{11}+n_{10}+n_{01}}{2N}\right)\left(\frac{2n_{11}+n_{10}+n_{01}-1}{2N-1}\right)+\left(\frac{2n_{00}+n_{01}+n_{10}}{2N}\right)\left(\frac{2n_{00}+n_{01}+n_{10}-1}{2N-1}\right).
\end{equation*}
Gwet in his book \parencite{gwet2014handbook} pointed out an alternative angle to see how $\alpha$ was defined. He narrated that $\alpha$ was defined based on an alternative estimator of total agreement, which is given as $\hat{p}_{a|\alpha}=[1-1/(2N)](n_{11}+n_{00})/N+1/(2N)$, but it used the same chance agreement estimator as in Scott's $\pi$ ($\hat{p}_{e|\pi}$). No matter in which of these two ways the readers might comprehend Krippendorf's $\alpha$, the resulting $\alpha$ estimators will be the same.
 
\textcite{van2019new} proposed a general agreement coefficient that can fit several popular IRA measures (i.e., Scott's $\pi$, Krippendorf's $\alpha$, Bennett et al.'s $S$, Cohen's $\kappa$, and Gwet's $\text{AC}_1$) into the same equation framework. He invoked similar but more relaxed assumptions as in \textcite{gwet2008computing} and used a similar derivation process as that for the Perreault-Leigh coefficient \parencite{perreault1989reliability}. Furthermore, he proposed a generalized IRA measure $I_r^2$ by incorporating Bayesian estimates of category proportions. Based on a weakly informative prior belief, the estimated chance agreement in the proposed IRA estimate can be expressed as
\begin{equation*}    
\hat{p}_{e|I_r^2}=\left(\frac{2n_{11}+n_{10}+n_{01}+1}{2N+2}\right)^2+\left(\frac{2n_{00}+n_{01}+n_{10}+1}{2N+2}\right)^2.
\end{equation*}

Mak's $\tilde{\rho}$ \parencite{mak1988analysing} is an IRA method proposed for the dichotomous-response case with two or more raters. It is derived based on the sample estimate of the between-rater correlation of rating responses. Under the two-rater case, an interpretation of its chance-agreement component was given in \parencite{blackman1993estimating} as $1$ minus the difference between the probability that the two raters’ responses would differ for all rating subjects and the probability that the two's responses would be different for the same rating subject. Thus, its chance agreement estimator can be expressed as
\begin{align*}
\hat{p}_{e|\tilde\rho}
&=1-\frac{(2n_{11}+n_{10}+n_{01})(2n_{00}+n_{10}+n_{01})-(n_{10}+n_{01})}{2N(N-1)}.
\end{align*}
Given the mathematical forms of the chance agreement estimators, it is clear to see that $\hat{p}_{e|\alpha}$ and $\hat{p}_{e|I_r^2}$ are simple variants of $\hat{p}_{e|\pi}$, and the three will be approximately equal to each other when the sample size of rating subjects $N$ approaches infinity. Regarding the complete formulas of the IRA statistics that will be shown later, Krippendorff's $\alpha$, Van Oest's $I_r^2$, and Mak's $\tilde{\rho}$ are all asymptotically equivalent to Scott's $\pi$.

Cohen's $\kappa$, which remains the most popular IRA statistic in a variety of scientific disciplines, intends to improve Scott's $\pi$ by relaxing the homogeneity assumption of rating distributions. This well-known statistic was originally developed as an alternative for the $\chi^2$ statistic to quantify the categorical agreement rather than the categorical association \parencite{cohen1960coefficient}. It assumes that the two raters work independently, and they rate based on fixed but not necessarily the same marginal distributions. Similar to Scott's $\pi$, the chance agreement component of $\kappa$ is constructed by calculating the total probability of agreeing on each rating category based on the estimated marginal distribution of each rater and the assumption of independence. Therefore, its chance agreement estimator is given as
\begin{align*}
\hat{p}_{e|\kappa}
&=\left(\frac{n_{11}+n_{10}}{N}\right)\left(\frac{n_{11}+n_{01}}{N}\right)+\left(\frac{n_{00}+n_{01}}{N}\right)\left(\frac{n_{00}+n_{10}}{N}\right).
\end{align*}
If the estimated marginal distributions of the two raters are exactly the same, Cohen's $\kappa$ will then be equal to Scott's $\pi$. Reflected in the illustrative data, this suggests $\kappa=\pi$ when $n_{10}=n_{01}$.

Widely discussed in the literature \parencite{feinstein1990high, cicchetti1990high, xu2014interrater}, Cohen's $\kappa$ may produce some paradoxical or abnormal values compared to the impression given by the observed proportion of agreement when the underlying outcome (e.g., the disease to be diagnosed) is extremely rare or prevalent (or reflected as both raters' marginal probabilities for voting ``+'' are extremely high or low; known as ``prevalence effect''). Similar abnormalities also present when the marginal distributions of the two raters are too different (known as ``bias effect''). \textcite{cicchetti1990high} recommended reporting the average proportions of agreement in raters' ``positive'' and ``negative'' decisions, $\hat{p}_{\text{pos}}=2n_{11}/(2n_{11}+n_{10}+n_{01})$ and $\hat{p}_{\text{neg}}=2n_{00}/(2n_{00}+n_{01}+n_{10})$, as supplementary references for the interpretations of $\kappa$ and bias identifications. To tackle the aforementioned paradoxes (``prevalence effect'' and ``bias effect''), \textcite{byrt1993bias} proposed a prevalence-adjusted bias-adjusted kappa (PABAK), by replacing the diagonal counts as $n_{11}'=n_{00}'=(n_{11}+n_{00})/2$, replacing the off-diagonal counts as $n_{10}'=n_{01}'=(n_{10}+n_{01})/2$, and calculating an IRA measure based on the adjusted data through the original Cohen's $\kappa$ formula. In fact, the resulting expression will be exactly the same as that for a big family of equivalent IRA statistics, leading by Bennett et al.'s $S$ index. 

Bennett, Alpert, and Goldstein first named this category-dependent IRA statistic as $S$ in 1954 \parencite{bennett1954communications}. As reviewed in \textcite{zwick1988another} and \textcite{feng2013underlying}, some equivalent statistics include Holley and Guilford's $G$ \parencite{holley1964note}, Maxwell's R.E. coefficient \parencite{maxwell1977coefficients}, Janson and Vegelius's $C$ \parencite{janson1979generalizations}, Brennan and Prediger's $\kappa_n$ \parencite{brennan1981coefficient}, and others \parencite{finn1970note, potter1999rethinking}. The $S$ index follows the structure of \eqref{eq: cc-agree} and assumes pure random ratings uniformly distributed among rating categories. This suggests that the chance agreement is not related to rater or item characteristics but it equals to the reciprocal of the number of rating categories. Specifically, the chance agreement estimator in the dichotomous ratings is
\begin{equation*}
\hat{p}_{e|S}=\frac{1}{2}.
\end{equation*}
It is not hard to see that $\kappa=\pi=S$ when $n_{10}=n_{01}$ and $n_{11}=n_{00}$ in Table \ref{tb: 2x2tb}. Also, it has been shown that $S\ge \kappa$ always holds \parencite{feinstein1990high}. On the other hand, some criticized that since $S=2\hat{p}_a-1$ is merely a simple linear transformation of $\hat{p}_a$, it includes no adjustment for chance \parencite{zwick1988another, byrt1993bias}.

There is another ``adjusted'' kappa-statistic proposed by \textcite{hoehler2000bias}, which exploits the idea of constructing the Receiver Operating Characteristic curve and optimizing the sensitivity and specificity. In his ``adjusted'' kappa calculation, the adjustments made on the Table \ref{tb: 2x2tb} data are as follows: $n_{11}''=n_{00}''=N\sqrt{\text{OR}}/[2(1+\sqrt{\text{OR}})]$, where $\text{OR}=n_{11}n_{00}/(n_{10}n_{01})$ is the common odds ratio for $2\times 2$ table and $n_{10}''=n_{01}''=0.5N-n_{11}''$. Interestingly, \textcite{walter2001hoehler} showed that Hoehler's adjusted kappa score is equivalent to Yule's $Y$ in mathematics. Yule's $Y$ coefficient \parencite{yule1912methods}, often known as the ``coefficient of colligation,'' was originally a correlation method developed earlier than most measures in this review. However, there have been voices suggesting $Y$'s potential to measure IRA. \textcite{spitznagel1985proposed} indicated that $Y$ could serve as a good alternative for measuring agreement to help combat $\kappa$'s and similar methods' problem of strong sensitivity to extreme prevalence. \textcite{hoehler2000bias} also claimed via some numerical assessment that $Y$ is able to help cope with the two ``kappa paradoxes.'' Due to these important connections, we believe that it is meaningful to further systematically compare $Y$'s empirical performance with other chance-corrected IRA measures. Using the notations in Table \ref{tb: 2x2tb}, Yule's $Y$ can be estimated by
\begin{equation*}
Y=\frac{\sqrt{\text{OR}}-1}{\sqrt{\text{OR}}+1}=\frac{\sqrt{n_{11}n_{00}}-\sqrt{n_{10}n_{01}}}{\sqrt{n_{11}n_{00}}+\sqrt{n_{10}n_{01}}}.
\end{equation*}
Importantly, $Y$ is not defined when $n_{10}=0$ or $n_{01}=0$. Therefore, when estimating $Y$ in small-sample cases, a $0.5$-continuity correction was recommended.

Maxwell and Pilliner’s $r_{11}$ \parencite{maxwell1968deriving}, which was derived from psychometric theory \parencite{landis1975review} (some key derivations can be found in Web Appendix A), also incorporates a sense of chance removal. However, when organizing the formula of $r_{11}$ into the structure of \eqref{eq: cc-agree}, $\hat{p}_{e|r_{11}}$ cannot be simplified into an explainable form. The full $r_{11}$ estimator was proposed as
\begin{equation*}
r_{11}=\frac{2(n_{11}n_{00}-n_{10}n_{01})}{(n_{11}+n_{10})(n_{01}+n_{00})+(n_{11}+n_{01})(n_{10}+n_{00})}.
\end{equation*}
The $r_{11}$ estimator has a similar expression as $\kappa$'s, with slight difference in the denominator. Therefore, caution should be made when distinguishing these two statistics. 

As discussed in \textcite{bartko1966intraclass, landis1975review, fleiss1975measuring, blackman1993estimating}, under a perspective of mixed-effects model and the Analysis of Variances (ANOVA) framework, Scott’s $\pi$, Mak’s $\tilde{\rho}$, Cohen's $\kappa$, and Maxwell and Pilliner’s $r_{11}$ all can be formulated and interpreted as both chance-corrected IRA measures and intraclass correlation coefficients. In Web Appendix A, we provided a summary of the derivation approaches mentioned in the literature above, which help reveal the dual interpretations of the four IRA measures and their interrelationships. Specifically, it suggested that $\pi$ and $\tilde{\rho}$ were derived under the assumption of no ``rater effects'', while $\kappa$ and $r_{11}$ were from the models with ``rater effects'' ($\kappa$ assumes random raters but $r_{11}$ assumes raters sampled from a fixed pool; see details in Web Appendix A). This echoes $\pi$'s underlying assumption of rater homogeneity. %as well as the more flexible conditions for $\kappa$ (and $r_{11}$). 
And in terms of the differences in model constraints, it implies that $\kappa$, with more flexible assumptions of ``rater effects'', is a more conservative estimator of IRA compared to $\pi$ and $r_{11}$ \parencite{landis1975review}. In addition, some finite-sample relationships among these four statistics have been shown \parencite{fleiss1975measuring, blackman1993estimating},
\begin{align*}
    |r_{11}|\ge|\kappa|,~~\kappa\ge \pi,~~\tilde{\rho}\ge \pi,~~r_{11}\ge \pi,
\end{align*}
where the equalities will all hold when $n_{10}=n_{01}$ in Table \ref{tb: 2x2tb}. 

\textcite{gwet2008computing} assumed that raters would either feel confident (and make $100\%$-correct judgement) or unsure on rating questions, which is similar to the assumptions proposed in \textcite{grove1981reliability} and \textcite{aickin1990maximum}. He assumed that feeling uncertain would lead to random guessing, and the chance agreement would emerge when at least one rater judged with random guessing. By decomposing the probability of chance agreement into two parts and approximating ``the probability that at least one rater performed random rating'' with a normalized measure of randomness in the form of a variance ratio, Gwet derived a paradox-resistant IRA measure called $\text{AC}_1$ (see, more details in Chapter 5 of \textcite{gwet2014handbook}). In Zhao et al.'s reviews \parencite{zhao2012reliability, zhao2013assumptions}, $\text{AC}_1$ was categorized into a separate class among the chance-corrected agreement measures that differed from the ``category-based'' Bennett et al.'s $S$ and the equivalent as well as the ``distribution-based'' IRA statistics including $\pi$, $\alpha$, $\kappa$, and their variants, and the author claimed that $\text{AC}_1$ could be regarded as ``double-based''. Compared with the chance-agreement estimator in Scott's $\pi$, its form in $\text{AC}_1$ has the ``positive'' distribution rate ($\hat{\omega}$) and the ``negative'' one ($1-\hat{\omega}$) switched. The corresponding chance-agreement component can be estimated by
\begin{align*}
    \hat{p}_{e|\text{AC}_1}
    &=2\hat{\omega}(1-\hat{\omega})\\
    &=2\left(\frac{2n_{11}+n_{10}+n_{01}}{2N}\right)\left(\frac{2n_{00}+n_{10}+n_{01}}{2N}\right).
\end{align*}
Interestingly, if trying to explain this equation in words, it suggests that the chance agreement component in $\text{AC}_1$ is estimated by the probability of making disagreement when the two raters independently vote following identical marginal distributions. Of note, when $\hat{\omega}=1/2$ or $n_{11}=n_{00}$, it satisfies $\text{AC}_1=\pi$.

\begin{table}[!th]
\caption{Point Estimations of Interrater Agreement (IRA) Statistics Under Two-Rater Dichotomous-Response Case}\label{tb: IRA-sum}
\begin{center}
\resizebox{\textwidth}{!}{
\begin{tabular}{ c c c }
\hline
IRA statistics & Estimator of chance agreement & Calculation formula \\ 
\hline
\multirow{3}{*}{Scott's $\pi$ (1955)} & &\\
& $\left(\frac{2n_{11}+n_{10}+n_{01}}{2N}\right)^2+\left(\frac{2n_{00}+n_{01}+n_{10}}{2N}\right)^2$ & $\frac{4(n_{11}n_{00}-n_{10}n_{01})-(n_{10}-n_{01})^2}{(2n_{11}+n_{10}+n_{01})(2n_{00}+n_{01}+n_{10})}$ \\
& &\\
%\hline
\multirow{3}{*}{Krippendorff's $\alpha$ (1970)} & &\\
& $\left(\frac{2n_{11}+n_{10}+n_{01}}{2N}\right)\left(\frac{2n_{11}+n_{10}+n_{01}-1}{2N-1}\right)+\left(\frac{2n_{00}+n_{01}+n_{10}}{2N}\right)\left(\frac{2n_{00}+n_{01}+n_{10}-1}{2N-1}\right)$ & $1-\frac{(2N-1)(n_{10}+n_{01})}{(2n_{11}+n_{10}+n_{01})(2n_{00}+n_{01}+n_{10})}$ \\
& &\\
%\hline
\multirow{3}{*}{Van Oest's $I_r^2$ (2019)} & &\\
& $\left(\frac{2n_{11}+n_{10}+n_{01}+1}{2N+2}\right)^2+\left(\frac{2n_{00}+n_{01}+n_{10}+1}{2N+2}\right)^2$ & $\frac{n_{11}+n_{00}-N\hat{p}_{e|I_r^2}}{N-N\hat{p}_{e|I_r^2}}$ \\
& &\\
%\hline
\multirow{3}{*}{Mak's $\tilde{\rho}$ (1988)} & &\\
& $1-\frac{(2n_{11}+n_{10}+n_{01})(2n_{00}+n_{01}+n_{10})-(n_{10}+n_{01})}{2N(N-1)}$ & $\frac{4(n_{11}n_{00}-n_{10}n_{01})-(n_{10}-n_{01})^2+(n_{10}+n_{01})}{(2n_{11}+n_{10}+n_{01})(2n_{00}+n_{01}+n_{10})-(n_{10}+n_{01})}$ \\
& &\\
\hline
%\hline
\multirow{3}{*}{Cohen's $\kappa$ (1960)} & &\\
& $\left(\frac{n_{11}+n_{10}}{N}\right)\left(\frac{n_{11}+n_{01}}{N}\right)+\left(\frac{n_{10}+n_{00}}{N}\right)\left(\frac{n_{01}+n_{00}}{N}\right)$ & $\frac{2(n_{11}n_{00}-n_{10}n_{01})}{(n_{11}+n_{10})(n_{10}+n_{00})+(n_{11}+n_{01})(n_{01}+n_{00})}$ \\
& &\\
%\hline
\multirow{3}{*}{Bennett et al.'s $S$ (1954)} & &\\
& $\frac{1}{2}$ & $\frac{2(n_{11}+n_{00})}{N}-1$ \\
& &\\
%\hline
\multirow{3}{*}{Yule 's $Y$ (1912)} & &\\
& $-$ & $\frac{\sqrt{n_{11}n_{00}}-\sqrt{n_{10}n_{01}}}{\sqrt{n_{11}n_{00}}+\sqrt{n_{10}n_{01}}}$ \\
& &\\
%\hline
\multirow{3}{*}{Maxwell and Pilliner's $r_{11}$ (1968)} & &\\
& $-$ & $\frac{2(n_{11}n_{00}-n_{10}n_{01})}{(n_{11}+n_{10})(n_{01}+n_{00})+(n_{11}+n_{01})(n_{10}+n_{00})}$ \\
& &\\
\hline
%\hline
\multirow{3}{*}{Gwet's $\text{AC}_1$ (2008)} & &\\
& $2\left(\frac{2n_{11}+n_{10}+n_{01}}{2N}\right)\left(\frac{2n_{00}+n_{01}+n_{10}}{2N}\right)$ & $\frac{2(n_{11}^2+n_{00}^2)-(n_{10}+n_{01})^2}{2(n_{11}N+n_{00}N-2n_{11}n_{00})+(n_{10}+n_{01})^2}$ \\
& &\\
\bottomrule
\end{tabular}}
\end{center}

\small\textit{Note}. Scott's $\pi$ (1955) is equivalent to Bryt et al.'s BAK (1993). Bennett et al.'s $S$ (1954) is equivalent to Bryt et al.'s PABAK (1993). Yule's $Y$ (1912) is equivalent to Hoehler's adjusted kappa (2000).
\end{table}

\begin{figure} [!th]
\centering
\includegraphics[width=1\textwidth]{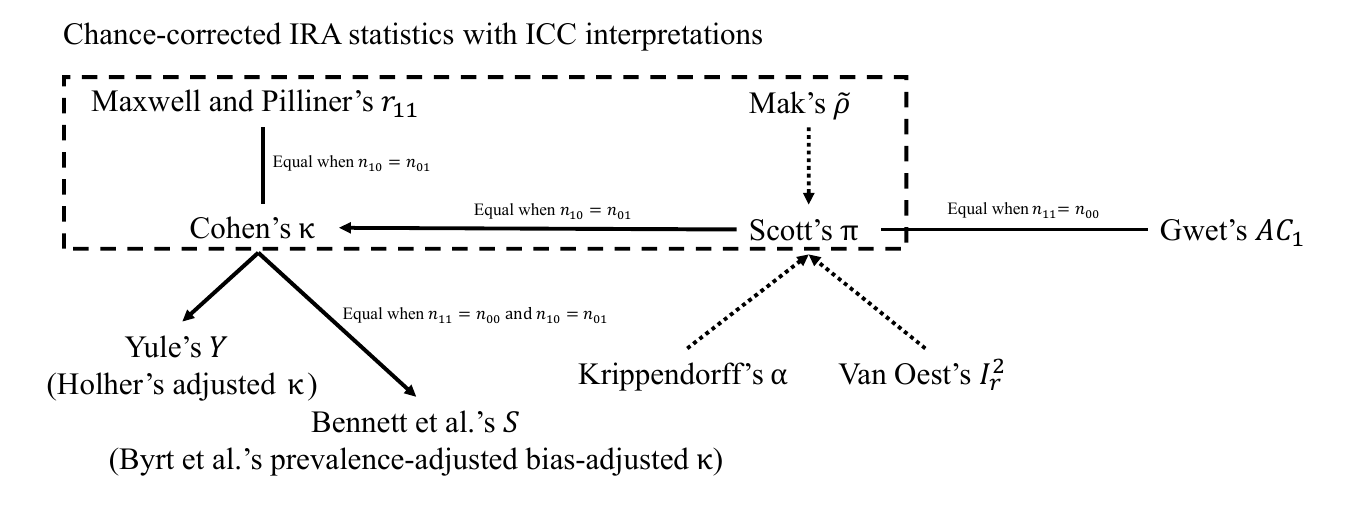}
\caption{Connections among the included interrater agreement (IRA) statistics. Solid lines with arrow represent method extension with additional considerations, solid lines without arrow indicate connection in mathematical form, and dashed lines with arrow suggest the directions of asymptotic convergence.} \label{fig: IRAS-web} 
\end{figure}

The chance agreement estimators (if applicable) and the complete calculation formulas of the IRA methods reviewed above are provided in Table \ref{tb: IRA-sum}. Finally, to illustrate the interrelationships among the reviewed IRA statistics, a diagram was provided to describe the mathematical and interpretation connections among their estimation formulas in Figure \ref{fig: IRAS-web}. %In fact, there are some other IRA statistics also incorporating the sense of chance-correction and popularly discussed in previous review papers \parencite{banerjee1999beyond}. We did not include them because they were not rigorously in the form of \eqref{eq: cc-agree}, nor did they have subtle relationships with other deemed chance-corrected IRA statistics, which might bring unfairness when conducting the comparisons in our subsequent simulation studies. For details about these methods, see, Perreault and Leigh's $I_r$ estimator \parencite{perreault1989reliability}, Zhao's $a_i$ estimator \parencite{zhao2012reliability}, and Pearson's tetrachoric correlation coefficient \parencite{pearson1901}.

\subsection{Variance estimation}
Variance provides a means of assessing and reporting the precision of a point estimate, %. It accounts for the uncertainty in the estimation of chance-corrected agreement and 
playing a critical role in the interval estimation, hypothesis testing, and power calculation. In this section, asymptotic variance formulas for different IRA statistics in the review are summarized, which allows approximate variance estimations for those IRA measures based on the observed data from Table \ref{tb: 2x2tb}. %The variance estimates will be valid in constructing confidence interval and making inferences about the agreement measures when the sample size of rating subjects are reasonably large.

\textcite{fleiss1969large} proposed a variance estimator for Cohen's $\kappa$ under the nonnull case of IRA, and \textcite{gwet2008computing} rewrote the variance formula under the two-rater case for better comparability with the estimated variances of several other IRA statistics. The estimated variance of $\kappa$ is given as
\begin{align*}
\widehat{\text{Var}}(\kappa)
&=\frac{1}{N(1-\hat{p}_{e|\kappa})^2}\Big\{ \hat{p}_a(1-\hat{p}_a)-4(1-\kappa) \left[ \hat{p}_{11}\hat{\omega}+\hat{p}_{00}(1-\hat{\omega})-\kappa\hat{p}_{e|\kappa}\right]\\
&+ 4(1-\kappa)^2\left[ \hat{p}_{11}\hat{\omega}^2 + \frac{1}{4}\hat{p}_{10}(2\hat{p}_{01}+\hat{p}_{a})^2 + \frac{1}{4}\hat{p}_{01}(2\hat{p}_{10}+\hat{p}_{a})^2 + \hat{p}_{00}(1-\hat{\omega})^2\right] \Big\},
\end{align*}
where $\hat{\omega}=\hat{p}_{11}+\hat{p}_{10}/2+\hat{p}_{01}/2$. The performance of this large-sample variance of $\kappa$ have been evaluated via Monte Carlo simulations in \textcite{cicchetti1977comparison} and \textcite{fleiss1978inference}. 

For Bennett et al.'s $S$ and the several equivalent IRA measures mentioned in the review, since the asymptotic variance estimator of percent agreement could be obtained as $\widehat{\text{Var}}(\hat{p}_a)=\hat{p}_a(1-\hat{p}_a)/N$ based on the binomial properties, we can get the variance estimator for $S=2\hat{p}_a-1$ as
\begin{align*}
\widehat{\text{Var}}(S)
=\frac{4}{N} \hat{p}_a(1-\hat{p}_a).
\end{align*}

For Scott's $\pi$, \textcite{gwet2008computing} proposed a nonparametric variance estimator with linearization techniques to remedy the formula proposed by \textcite{fleiss1971measuring}, which was under the hypothesis of no agreement and was not valid for building up confidence intervals. Under our scenario of interest, the variance formula of $\pi$ can be written as 
\begin{align*}
\widehat{\text{Var}}(\pi)
&=\frac{1}{N(1-\hat{p}_{e|\pi})^2}\Big\{ \hat{p}_a(1-\hat{p}_a)-4(1-\pi) \left[\hat{p}_{11}\hat{\omega}+\hat{p}_{00}(1-\hat{\omega})-\hat{p}_a\hat{p}_{e|\pi}\right]\\
&+ 4(1-\pi)^2\left[ \hat{p}_{11}\hat{\omega}^2+\frac{1}{4}(\hat{p}_{10}+\hat{p}_{01})+\hat{p}_{00}(1-\hat{\omega})^2-\hat{p}_{e|\pi}^2\right] \Big\}.
\end{align*}

Regarding Gwet's $\text{AC}_1$, \textcite{gwet2008computing} utilized a linear approximation that included all terms with a stochastic order of magnitude up to $N^{-1/2}$ to derive a consistent variance estimator. The variance formula of $\text{AC}_1$ is given by
\begin{align*}
\widehat{\text{Var}}(\text{AC}_1)
&=\frac{1}{N(1-\hat{p}_{e|\text{AC}_1})^2}\Big\{ \hat{p}_a(1-\hat{p}_a)-4(1-\text{AC}_1) \left[\hat{p}_{11}(1-\hat{\omega})+\hat{p}_{00}\hat{\omega}-\hat{p}_a\hat{p}_{e|\text{AC}_1}\right]\\
&+ 4(1-\text{AC}_1)^2\left[ \hat{p}_{11}(1-\hat{\omega})^2 + \frac{1}{4}(\hat{p}_{10}+\hat{p}_{01}) + \hat{p}_{00}\hat{\omega}^2-\hat{p}_{e|\text{AC}_1}^2\right] \Big\}.
\end{align*}
The variance estimator for $\pi$ is similar to that for $\text{AC}_1$ in the general structure, which also resonates their similar but ``reversed'' structures of the chance-agreement estimator.

In \textcite{gwet2014handbook}, the author used similar approaches to derive a general variance formula for Krippendorff's $\alpha$. Under the $2\times 2$ case of interest, the variance estimator is shown as follows,
\begin{align*}
    \widehat{\text{Var}}(\alpha)
    &=\frac{1}{N(1-\hat{p}_{e|\alpha}')^2}\Big\{\hat{p}_{11}(1-\hat{\omega}+\hat{\omega}\alpha')^2+\frac{1}{4}(\hat{p}_{10}+\hat{p}_{01})(1-\alpha')^2+\hat{p}_{00}(\alpha'+\hat{\omega}-\hat{\omega}\alpha')^2\\
    &-[\alpha'-\hat{p}_{e|\alpha}'(1-\alpha')]^2\Big\},
\end{align*}
where $\hat{p}_{e|\alpha}'=\hat{p}_{e|\pi}$ and $\alpha'=\pi$, again, suggesting $\alpha$'s close relationship with Scott's $\pi$. 

\textcite{bloch19892} derived the asymptotic variance for the maximum likelihood estimator of intraclass kappa $\hat{\kappa}_I$, which is an IRA estimator equivalent to Scott's $\pi$. The large-sample variance formula can be given as
\begin{align*}
    \widehat{\text{Var}}(
    \hat{\kappa}_I)=\frac{1-\hat{\kappa}_I}{N}\left[(1-\hat{\kappa}_I)(1-2\hat{\kappa}_I)+\frac{\hat{\kappa}_I(2-\hat{\kappa}_I)}{2\hat{\omega}(1-\hat{\omega})}\right].
\end{align*}
Recall that \textcite{blackman1993estimating} have proved the asymptotic equivalence among the estimators $\kappa$, $\pi$, $\tilde{\rho}$, and $r_{11}$, we have also shown the large-sample equivalence of Scott's $\pi$ and Van Oest's $I_r^2$ in the method review. We found that the derivations of the large-sample variance estimators discussed in \textcite{bloch19892} will yield the same results for $\tilde{\rho}$, $r_{11}$, and $I_r^2$. In other words, we can use the asymptotic variance formula above also for $\tilde{\rho}$, $r_{11}$, and $I_r^2$ after appropriate notation substitutions at the position of $\hat{\kappa}_I$.

Assuming multinomial-distributed counts in Table \ref{tb: 2x2tb} and applying the delta method, the asymptotic variance estimator for Yule's $Y$ can be derived as
\begin{align*}
    \widehat{\text{Var}}(Y)=\frac{1}{16}(1-Y^{2})^2\left(\frac{1}{n_{11}}+\frac{1}{n_{10}}+\frac{1}{n_{01}}+\frac{1}{n_{00}}\right).
\end{align*}
Based on \textcite{agresti2003categorical} and \textcite{bonett2007statistical}'s recommendation, a $0.1$- or $0.5$-continuity correction should be added to each observed count for bias reduction. Furthermore, \textcite{bonett2007statistical} suggested a refined $100(1-\alpha)$ percent approximate confidence interval for $Y$ based on Fisher's $z$-transformation, which was shown to have much better small-sample properties compared to the Wald confidence interval. With our notations, the refined large-sample confidence interval is constructed as
\begin{equation*}
    \tanh\left[\tanh^{-1}(Y)\pm z_{1-\frac{\alpha}{2}}\frac{1}{4}\sqrt{\frac{1}{n_{11}+0.5}+\frac{1}{n_{10}+0.5}+\frac{1}{n_{01}+0.5}+\frac{1}{n_{00}+0.5}}\right],
\end{equation*}
where $z_{1-\alpha/2}$ is the $1-\alpha/2$ quantile of the standard normal distribution.

\section{A generalized IRA framework with correlated decisions} \label{sec: 3}
In this section, we propose a new data-generating framework to depict the uncertainty involved in a rating process. It expands the IRA frameworks in the literature by introducing correlated decisions in the rating process between two raters.

In many previous simulation-based comparative studies, the two raters' decisions were assumed to be completely independent \parencite{grant2017evaluation, xu2014interrater, gwet2008computing, feng2013factors, feng2013underlying, sym14020262, van2019new}. Because the two raters make judgments on the same set of rating subjects and based on the same judgment criteria or similar training backgrounds, their responding behaviors on those subjects might be correlated. For example, it is likely that the two raters both encounter difficulty about the intractable tumor image from specific subjects, or the two might simultaneously make false decisions against the unobserved truth if the rating subject exhibited some delusive elements. The correlation could even be strong when the two raters' training backgrounds get close, or the subjects possess significant features. As a result, failing to simulate these correlated features in rater decisions might largely distort the assessment of IRA statistics. 

To address this potential issue, here we develop a generalized data-generating framework to mimic more realistic two-rater dichotomous rating tasks. We still assume that the raters make independent ratings, but in a sense that their decisions are conditionally independent to the rater's training background, the rating guidelines, or specific features in each rating subject. On the other hand, we do not attempt to characterize these features directly to avoid the overwhelming complexity. Instead, these unobserved features introduce, mathematically, a marginal between-rater correlation. It determines the overall degree of \textit{probabilistic certainty} within total uncertainty for agreement, which we desire to capture in the IRA statistics.  To evaluate the performance of IRA statistics, we also define a new measure for the ``truth'' of chance-corrected IRA, consistent with the development framework, as the benchmark. It can overcome the potential (under-estimation) bias by ignoring between-rater correlation, to facilitate the assessment of IRA statistics in the simulation studies in Section \ref{sec: 4}. 

\subsection{Assumptions and notations}

\begin{figure}[!t]
\centering
\includegraphics[scale=1.2]{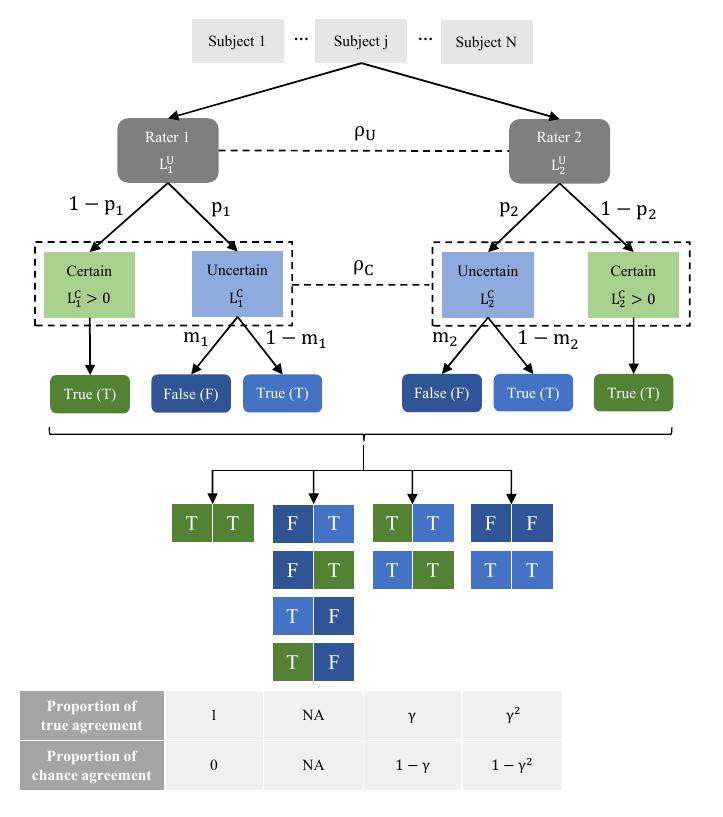}
\caption{Illustrations of the data generation and agreement-type specifications in the proposed simulation framework. $N$ is the total sample size of rating subjects, $\theta$ is the prevalence of the positive outcome, $p_i$ is the probability to encounter uncertainty for Rater $i$, $i=1,2$, $m_i$ is the probability to make a wrong decision against the ``true'' outcome, $L_i^U$ is the latent variable behind $p_i$, $L_i^C$ is the latent variable behind $m_i$, $\rho_U$ is the correlation between $L_1^C$ and $L_2^C$, and $\rho_C$ is the correlation between $L_1^U$ and $L_2^U$. 
} \label{fig: notations} 
\end{figure}

Assume that Rater $1$ and Rater $2$ make binary votes, positive ($+$) or negative ($-$), on the same set of subjects with a total sample size $N$ (Figure \ref{fig: notations}). For each of the rating subjects, we impose a two-step decision process: the two raters may feel \textit{certain} or \textit{uncertain} about each rating subject in step I, and make correct decisions in probability 1 (for \textit{certain} cases) or with informative guessing (for \textit{uncertain} cases) in step II. As mentioned earlier, the rater's decisions are conditionally independent in our setting, which includes the complete independence assumption (in other's work) as a special case with correlation coefficients fixed at $0$.

To apply the two-step decision process to generate raters’ voting, we first generate the underground truth of each rating subject $j$, $j=1,..., N$. The underlying outcome prevalence, or the true proportion of ``+'' outcomes in the population of rating subjects, is denoted as $\theta$, which is a fixed attribute independent of specific rater's response. In our simulation process, we generate raters' votes by their consistency with respect to the underlying subject truth (i.e., ``True (T)'' vs ``False (F)'' in Figure \ref{fig: notations}), which can be later converted into votes (i.e., ``$+$'' vs ``$-$'' ) and aggregated into a $2\times2$ table (e.g., Table \ref{tb: 2x2tb}). In step I of the two-step decision process, we use a rater-specific probability $p_i$ to denote the probability to encounter uncertainty for Rater $i$, $i=1,2$. 
In step II, \textit{correct} decisions aligned with the underlying truth are always made in the \textit{certain} cases; for the \textit{uncertain} cases, a rater's subject-specific decision is exposed to the risk of misclassification, with a probability $m_i$ to make a wrong decision opposite to the underlying truth. Small $m_i$ indicates high accuracy of guessing from Rater $i$, which also partially describes the rater's characteristics (e.g., a professional radiologist could still make a good judgment whether the subject is actually a disease case even though the evidence may not be completely deterministic).

In step I, we incorporate the underlying correlation of rater behaviors using two continuous latent variables $(L_{1j}^U, L_{2j}^U)$. By introducing correlated latent variables, the difficulty of directly specifying correlation between binary variables could be coped with \parencite{albert1993bayesian}. Detailed justifications of these latent variables are provided in Web Appendix B. In brief, $(L_{1j}^U, L_{2j}^U)$ denotes the Rater $1$ and $2$'s bivariate latent variables that controls the raters' binary uncertainty status on rating subject $j$, which is determined by the threshold 0 --- $L_{ij}^U\le 0$ indicates Rater $i$ feels \textit{certain} on rating subject $j$, while $L_{ij}^U>0$ indicates feeling \textit{uncertain} in decision. Additionally, we use a correlation coefficient $\rho_U$ to quantify the correlation between the two raters' latent variables $(L_{1j}^U, L_{2j}^U)$. We practically assume $\rho_U\geq 0$, suggesting the raters may somehow simultaneously encounter difficulties and feel \textit{uncertain} due to the same set of working subject and potentially similar training. 

In step II, we exploit two more bivariate latent variables $(L_{1j}^C, L_{2j}^C)$ to characterize whether each rater makes the correct decision or not. Another correlation coefficient $\rho_C \geq 0$ controls the degree of correlation between the two raters' latent variables $(L_{1j}^C, L_{2j}^C)$. The sign of $L_{ij}^C$ indicates whether Rater $i$ makes a decision consistent with the truth of subject $j$. For the \textit{certain} subjects, Rater $i$ will have $L_{ij}^C$ follow a truncated normal distribution (such that $L_{ij}^C>0$), indicating an always correct decision. For rating subjects that one rater is \textit{certain} and another is \textit{uncertain}, the latent variable $L_{ij}^C$ of the \textit{uncertain} rater follows a conditional normal distribution, and the binary vote can be determined accordingly given $L_{ij}^C$.

A step-by-step description of our data-generating mechanism and the related mathematical details are provided in Web Appendix C. And, an illustrative flowchart is displayed in Figure \ref{fig: notations} to sketch the general structure of our proposed data-generating framework with correlated ratings. Based upon this data-generation framework, we can simulate the observed rating responses given by the two correlated raters, which can be finally tabulated into a $2\times 2$ table-formatted data for IRA statistic calculations and subsequent comparisons.

\subsection{``Truth'' table and rater agreement}
Following the chance-correction definition in \eqref{eq: cc-agree}, we propose an estimand of ``true'' chance-corrected IRA by first expressing the true total agreement ($p_a$) and chance agreement ($p_e$) as mathematical functions of the $7$ key parameters defined in the data-generating framework --- the outcome prevalence ($\theta$), the two raters' probabilities of encountering uncertainty ($p_1$ and $p_2$), the two raters' probabilities of mistaken decisions ($m_1$ and $m_2$), the latent correlation measure of encountering uncertainty ($\rho_U$), and the latent correlation measure of making correct decisions ($\rho_C$). 

To scrutinize the constitution of $p_e$, we construct an ancillary $4\times 4$ ``truth'' table to classify the unobserved uncertainty statuses and the observed final decisions (Table \ref{tb: 4x4truth}). We denote $U_{..}$ the joint probabilities regarding the two raters' uncertainty status (the first and the second subscripts respectively correspond to Rater 1 and 2, and value $1$ means uncertainty while $0$ means certainty in voting), which are functions of $p_1$, $p_2$, and $\rho_U$; we denote $C_{..}$ the joint probabilities regarding the two raters' correct-decision status ($1$ means making correct decisions while $0$ means misclassifications), which are functions of $m_1$, $m_2$, and $\rho_C$; $C_{b|a}$ are the conditional probabilities of Rater $b$ making a correct decision given Rater $a$ feels \textit{certain}, where $a$ and $b$ are distinct rater indicators. In summary, each of the $4\times 4$ cells in Table \ref{tb: 4x4truth} is filled with the theoretical probability of certain rating result combination (e.g., $(+,+)$) conditional on certain unobserved uncertainty status combination (e.g., (Uncertain, Uncertain)), expressed with the $7$ key parameters. More details about the full mathematical expressions of the notations above and the related clarifications can be found in Web Appendix D. 

\begin{table}[htbp]
\caption{$4\times 4$ ``Truth'' Table Cross-Classified by Underlying Uncertainty Statuses and Observed Rating Responses}\label{tb: 4x4truth}
\begin{center}
\resizebox{\textwidth}{!}{\begin{tabular}{ c c c c c c c c }
\hline
 & & & \multicolumn{5}{c}{Rater 2}\\ \cline{4-8}
%\hline
 & & & \multicolumn{2}{c}{Uncertain} & & \multicolumn{2}{c}{Certain} \\ \cline{4-5} \cline{7-8}
%\hline
 & & & $+$ & $-$ & & $+$ & $-$\\
\hline
\multirow{10}{*}{Rater 1} & \multirow{5}{*}{Uncertain} & & & & & & \\
& & $+$ & $U_{11}[\theta C_{11}+(1-\theta)C_{00}]$ & $U_{11}[\theta C_{10}+(1-\theta)C_{01}]$ & & $U_{10}\theta C_{1|2}$ & $U_{10}(1-\theta)(1-C_{1|2})$\\
& &  &  &  &  & &\\
& & $-$ & $U_{11}[\theta C_{01}+(1-\theta)C_{10}]$ & $U_{11}[\theta C_{00}+(1-\theta)C_{11}]$ & & $U_{10}\theta(1- C_{1|2})$ & $U_{10}(1-\theta)C_{1|2}$\\
& &  &  &  &  & \\ 
%\hline
& \multirow{5}{*}{Certain} & & & & & &\\
& & $+$ & $U_{01}\theta C_{2|1}$ & $U_{01}\theta (1-C_{2|1})$ & & $U_{00}\theta$ & 0\\
& &  &  &  &  & &\\
& & $-$ & $U_{01}(1-\theta)(1- C_{2|1})$ & $U_{01}(1-\theta) C_{2|1}$ & & 0 & $U_{00}(1-\theta)$\\
& &  &  &  &  & &\\
\hline
\end{tabular}}
\end{center}
\small\textit{Note}. $U_{..}$ are the joint probabilities regarding the two raters' uncertainty status (the first and the second subscripts respectively correspond to Rater 1 and 2, and value $1$ means uncertainty while $0$ means certainty in voting), which are functions of $p_1$, $p_2$, and $\rho_U$. $C_{..}$ are the joint probabilities regarding the two raters' correct-decision status ($1$ means making correct decisions while $0$ means misclassifications), which are functions of $m_1$, $m_2$, and $\rho_C$. $C_{b|a}$ are the conditional probabilities of Rater $b$ making a correct decision given Rater $a$ feels \textit{certain}.
\end{table}

Using Table \ref{tb: 4x4truth}, the total agreement proportion $p_a$ can be obtained by summing up the diagonal cells, which can be simplified as
$$p_a=U_{11}(C_{11}+C_{00})+U_{10}C_{1|2}+U_{01}C_{2|1}+U_{00}.$$
Based on our assumptions in the previous subsection, ``chance agreement'' may emerge only when at least one rater feels \textit{uncertain}. When both raters feel \textit{certain} (the bottom-right $2\times 2$ block in Table \ref{tb: 4x4truth}), the agreement achieved is completely true without chance agreement. In the rest of three $2\times 2$ blocks when voting may be \textit{uncertain}, it is critical to quantify the overall degree of \textit{probabilistic certainty}, which should be distinguished from chance agreement caused by random voting. The \textit{probabilistic certainty} is controlled by the between-rater correlation ($\rho_C$). A $\rho_C \approx 1$ indicates that the raters tend to simultaneously make correct or wrong decisions. In other words, the raters' voting could be almost systematically identical for every subject even though the raters could be \textit{uncertain}, suggesting little chance agreement caused by random voting. On the contrary, $\rho_C = 0$ degenerates to the special case that the two raters independently make decisions. In this case, all the agreement made when at least one rater feels \textit{uncertain} is due to complete randomness and should be classified as chance agreement. Subsequently, the total agreement in the upper-left three blocks in Table \ref{tb: 4x4truth} can be partitioned into two components --- a probabilistic ``true agreement'' that is induced by that rater correlation $\rho_C$ and a remaining proportion of chance agreement. To identify the amount of ``true agreement'' under the uncertainty cases, we denote the proportion of ``true agreement'' (or \textit{probabilistic certainty}) by $\gamma_1$ for the case when only one rater feels \textit{uncertain} and by $\gamma_2$ for the case of both raters feeling \textit{uncertain}. Then, the ``true agreement'' can be expressed as
\begin{equation*}
    p_a-p_e = U_{00}+\gamma_1(U_{10}C_{1|2}+U_{01}C_{2|1})+\gamma_2 U_{11}(C_{11}+C_{00}).
\end{equation*} 
To quantify $\gamma_1$ and $\gamma_2$ with the parameters in hand, we consider using the correlation between the two raters' correct votes to do reasonable approximations. For bivariate binary outcomes, \textcite{marshall1985family} suggested using the Phi coefficient as a natural estimate of the correlation between binary variables in analogy with the Pearson's correlation coefficient, which can be written as
\begin{equation*}\gamma=\gamma(m_1,m_2,\rho_C)=\frac{C_{00}C_{11}-C_{10}C_{01}}{\sqrt{(C_{00}+C_{01})(C_{00}+C_{10})(C_{11}+C_{01})(C_{11}+C_{10})}}.
\end{equation*} 
Here, the correlation coefficient $\gamma$ indexes the direction and extent to which two raters are related. It reflects the tendency of the raters' voting to ``co-vary''; that is, for changes in the value of one voting to be associated with changes in the value of the other. Thus, we consider $\gamma$ as a surrogate for $\gamma_1$ under the case that one rater feels \textit{uncertain} and the other rater feels \textit{certain} and always make correct decisions, playing the role of the proportion of probabilistic true agreement due to the imposed correlated decisions. For the case that both raters feel \textit{uncertain}, we consider $\gamma^2$ as a surrogate for $\gamma_2$ in analogy with the R-square in simple linear regression. Finally, it enables us to provide an expression of the ``true'' chance-corrected IRA as follows,
\begin{align*}
    K=\frac{p_a-p_e}{1-p_e}=\frac{U_{00}+\gamma[\gamma U_{11}(C_{11}+C_{00})+U_{10}C_{1|2}+U_{01}C_{2|1}]}{1-(1-\gamma)[(1+\gamma)U_{11}(C_{11}+C_{00})+U_{10}C_{1|2}+U_{01}C_{2|1}]},
\end{align*}
which will be used for IRA assessment in the simulation study. In fact, the estimand $K$ does not rely on the prevalence $\theta$; hence it is attractive for serving as a benchmark of IRA assessments.

\section{Simulation studies}\label{sec: 4}
We conducted comprehensive simulations to assess the performances of a series of IRA statistics compared to the ``true'' measure $K$ defined above. 9 IRA statistics reviewed above as well as the observed proportion of agreement ($\hat{p}_a$) were considered in the evaluations.

Before introducing the simulation settings, we clarify an important fact behind our data-generating mechanism. Table $\ref{tb: 4x4truth}$ can imply the theoretical probabilities within the simulated $2\times 2$ tables given different choices of the $7$ key parameters. It follows
\begin{align*}
    p_{11}
    &=U_{11}(\theta C_{11}+(1-\theta)C_{00}) + U_{10}\theta C_{1|2} + U_{01}\theta C_{2|1} + U_{00}\theta, \\
    p_{10}
    &=U_{11}(\theta C_{10}+(1-\theta)C_{01}) + U_{10}(1-\theta)(1-C_{1|2}) + U_{01}\theta(1-C_{2|1}),\\
    p_{01}
    &=U_{11}(\theta C_{01}+(1-\theta)C_{10}) + U_{10}\theta(1-C_{1|2}) + U_{01}(1-\theta)(1-C_{2|1}),\\  
    p_{00}
    &=U_{11}(\theta C_{00}+(1-\theta)C_{11}) + U_{10}(1-\theta)C_{1|2} + U_{01}(1-\theta)C_{2|1} + U_{00}(1-\theta).
\end{align*}
It is straightforward to see that if replacing $\theta$ with $(1-\theta)$, then the expressions of $p_{11}$ and $p_{00}$ will change with each other while $p_{10}$ and $p_{01}$ will also interchange their values with other key parameters fixed. After this inversion of the outcome prevalence level, the theoretical values of all our included IRA measures will remain the same due to the symmetric features in their estimand expressions. This further suggests that the empirical performance of all these measures under the $\theta$ prevalence level is expected to be close to that under the $(1-\theta)$ prevalence setting given some extra sampling error. %In terms of this feature, we only show the simulation results for $\theta$ lower than 0.5 (i.e., extreme prevalence to middle-level prevalence), as those for $(1-\theta)$ have been validated to be nearly symmetric based on some extra simulations not included in this manuscript.

\subsection{Simulation settings and evaluation methods}
Based on our simulation framework, the benchmark IRA $K$ and the data generation are determined by the following parameters: the prevalence of the rating outcome ($\theta$), the two raters' marginal probabilities of feeling uncertain about an item ($p_1$ and $p_2$), the two raters' indirect correlation measure of encountering uncertainties ($\rho_U$), the two raters' marginal probabilities of giving a wrong judgment when uncertain ($m_1$ and $m_2$), the indirect correlation measure of making a correct decision ($\rho_C$), and the total sample size of rating subjects ($N$). In this assessment, we considered 9 levels of prevalence $\theta\in\{0.1, 0.2, ..., 0.9\}$, 5 levels of probability of feeling \textit{uncertain} for each rater $p_1, p_2\in\{0.1, 0.3, 0.5, 0.7, 0.9\}$, 5 levels of latent correlation parameter about uncertainty $\rho_U\in\{0.1, 0.3, 0.5, 0.7, 0.9\}$, 5 levels of misclassification rates given uncertainty $m_1, m_2\in\{0.1, 0.2, 0.3, 0.4, 0.5\}$ (we assume that rational decisions tend to perform not worse than random guessing, i.e., $m_i=0.5$), 5 levels of latent correlation parameter about making correct decisions $\rho_C\in\{0.1, 0.3, 0.5, 0.7, 0.9\}$, and 4 levels of total numbers of rating subjects $N\in \{25,50,100,200\}$.

Given each parameter combination, our proposed framework allows us to calculate a ``true'' IRA $K$ under this specific setting. Then, we generate $1,000$ $2\times 2$ tables following the proposed data-generating process and calculate the corresponding $10$ IRA measures under each data simulation. The bias calculated from the Monte Carlo mean of each IRA measure over $1,000$ simulated data minus the setting-specific $K$ value is of primary interest. The coverage probability of the ``true'' chance-corrected IRA, which is defined as the probability that $K$ fell into the $95\%$ confidence interval of each IRA measure over $1,000$ replications, is also computed for each IRA measure under each parameter setting. This comparison metric is for assessing how well the IRA measures' interval estimates can help to describe the benchmark $K$, and in the coverage calculations, the confidence intervals are constructed based on the asymptotic variance estimation methods included in the review. In addition, we also conducted an agglomerative hierarchical clustering using our extensive simulation data to estimate the similarity among the reviewed methods as well as our proposed ``true'' chance-corrected IRA $K$. Euclidean distance across different simulation settings and the average linkage method were used to quantify the between-IRA method distances and to build up a dendrogram. Our simulations were carried out in R (version 4.2).

\subsection{Simulation results}

\subsubsection{Overall evaluations across all settings}

\begin{figure}[htbp!]
\centering
\includegraphics[width=1.1\textwidth]{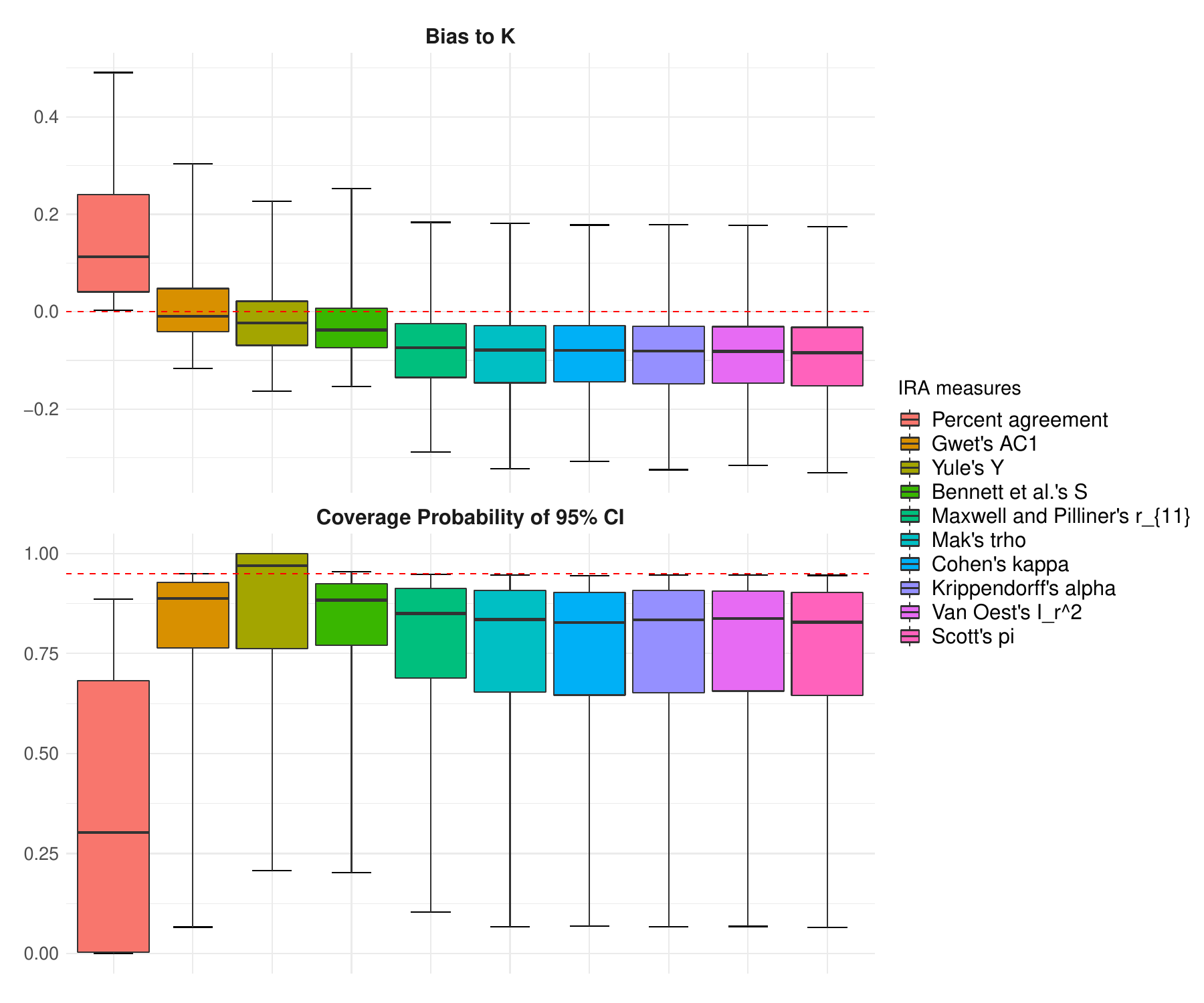}
\caption{Across-scenario overall bias towards the ``true'' interrater agreement $K$ and overall coverage probability using $95\%$ confidence intervals for $10$ interrater agreement measures over all $562,500$ simulation settings. From the bottom to the top, the five summary statistics in the box plots are the 2.5 percentile, the first quartile, the median, the third quartile, and the 97.5 percentile.} \label{fig: overall} 
\end{figure}

Figure \ref{fig: overall} shows the biases and coverage probabilities of different IRA measures compared to the ``true'' chance-corrected IRA $K$ among all the $9\times 5\times 5\times 5\times 5\times 5\times 5\times 4=562,500$ parameter constellations in our simulation study. In the boxplots, each IRA measure's mean bias or coverage rate over $1,000$ iterations under a single simulation setting was regarded as a data point. From left to right, the $10$ IRA measures were ranked based on the magnitudes of median bias values, and we kept this order through all the result presentations for consistency. These figures give an overall impression of how different IRA measures perform when no information about the raters or the rating task is acquired. Over the vast range of settings, Gwet's $\text{AC}_1$ (median bias$=-0.009$), Yule's $Y$ ($-0.023$), and Bennett et al.'s $S$ ($-0.038$) show smaller median bias across scenarios compared to other IRA methods. The overall bias performances of Maxwell and Pilliner's $r_{11}$, Mak's $\tilde{\rho}$, Cohen's $\kappa$, Van Oest's $I_r^2$, and Scott's $\pi$ are close to each other (medians vary from $-0.074$ to $-0.084$), all of which overall underestimate the ``true'' IRA $K$. Finally, the observed proportion of agreement $\hat{p}_a$, without any correction of chance agreement, will always overestimate $K$. Regarding the overall performance of each IRA measure's interval estimate, the coverage probabilities for $\text{AC}_1$, $Y$, and $S$ will be larger than $75\%$ under most settings. While the interval estimate for $Y$ seems to be too conservative in some scenarios (median coverage probability=$96.9\%$), $\text{AC}_1$ ($88.8\%$) and $S$ ($88.3\%$) show similar coverage performances across various simulation settings.

\begin{figure}[htbp!]
\centering
\includegraphics[width=1\textwidth]{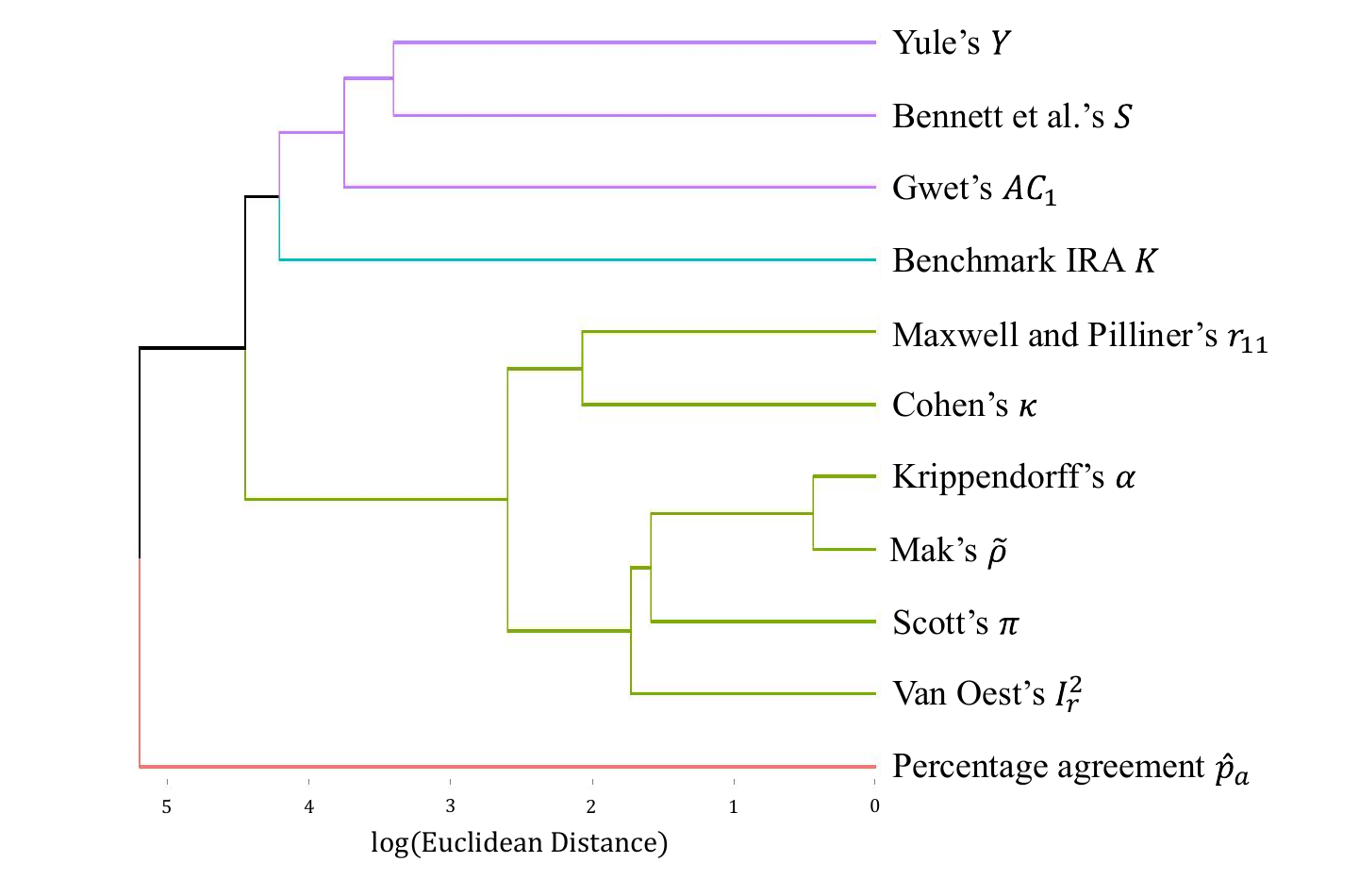}
\caption{Agglomerative hierarchical clustering of the $10$ IRA measures and the ``true'' interrater agreement $K$ based on $562,500$ simulation settings.} \label{fig: clustering} 
\end{figure}

Figure \ref{fig: clustering} gives an overall assessment about the closeness among the $10$ interrater agreement methods as well as the benchmark measure $K$ based on their estimates/values across all simulation settings. IRA methods with similar estimates across different scenarios are agglomerated into clusters. By gauging the change of within-cluster and between-cluster variabilities along the increase of cluster numbers, 3 or 4 clusters should be an optimal cutting number. The hierarchical clustering indicates that the percent agreement $\hat{p}_a$ itself forms a cluster, which highlights its different nature compared to other chance-corrected IRA methods. If classifying into 4 clusters, $K$ forms a cluster on itself, while if dividing into 3 clusters, $K$ will be combined into the cluster that includes $Y$, $S$, and $\text{AC}_1$. This, again, shows the relative closeness between $K$ and the three methods having shown advantages in the across-scenario bias assessment. Finally, $r_{11}$, $\kappa$, $\alpha$, $\tilde{\rho}$, $\pi$, and $I_r^2$ agglomerate a large cluster, suggesting their overall similarity across all simulation settings.

\subsubsection{Bias evaluations at fixed levels of key factors}
\begin{figure}[htbp!]
\centering
\includegraphics[width=1\textwidth]{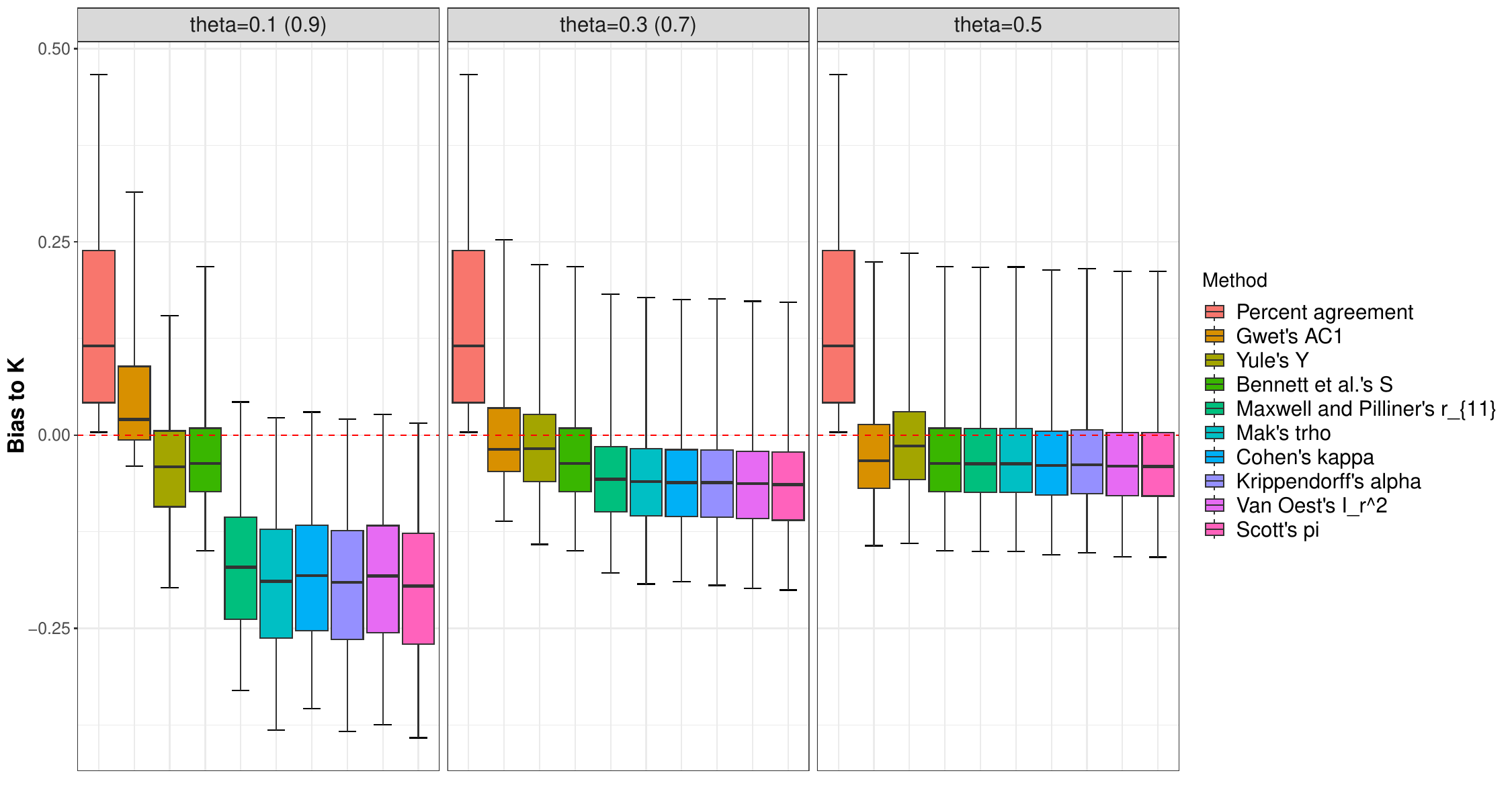}
\caption{Across-scenario overall bias to ``true'' interrater agreement $K$ for $10$ interrater agreement measures at different outcome prevalence levels.} \label{fig: theta_bias} 
\end{figure}

In this part, we explore the bias performance of different IRA methods compared to $K$ when fixing one or more of the key simulation parameters at specific levels that might have practical implications. Figure \ref{fig: theta_bias} shows the overall comparison at extreme to non-extreme prevalence levels (as reasoned above, the performances of all IRA statistics are approximately symmetric with respect to the $0.5$ prevalence, based on their symmetric estimand designs). From different panels, it shows that $S$ and $\hat{p}_a$ are invariant to $\theta$. When $\theta$ is extremely low or high, $\text{AC}_1$ performs the best on the aspect of median bias, followed by $S$ and $Y$, while the group of methods from $r_{11}$ to $\pi$ all show relatively large negative differences compared to the benchmark $K$. When $\theta$ approaches $0.5$, the group of similarly-performed methods and $\text{AC}_1$ tend to converge to a level close to the bias performance of $S$, and the underestimation pattern of that group of methods are gradually ameliorated when $\theta$ becomes non-extreme. On the other hand, at $\theta=0.5$, $Y$ becomes slightly better than all other methods regarding the bias to $K$.

Since it has been shown that the prevalence $\theta$ has a significant impact on differentiating various IRA statistics, and in practice, the extremity of prevalence level is usually discernible, we show the following results stratified by extreme or close-to-0.5-level prevalences to facilitate practical interpretations.

\begin{figure}[htbp!]
\centering
\includegraphics[width=1\textwidth]{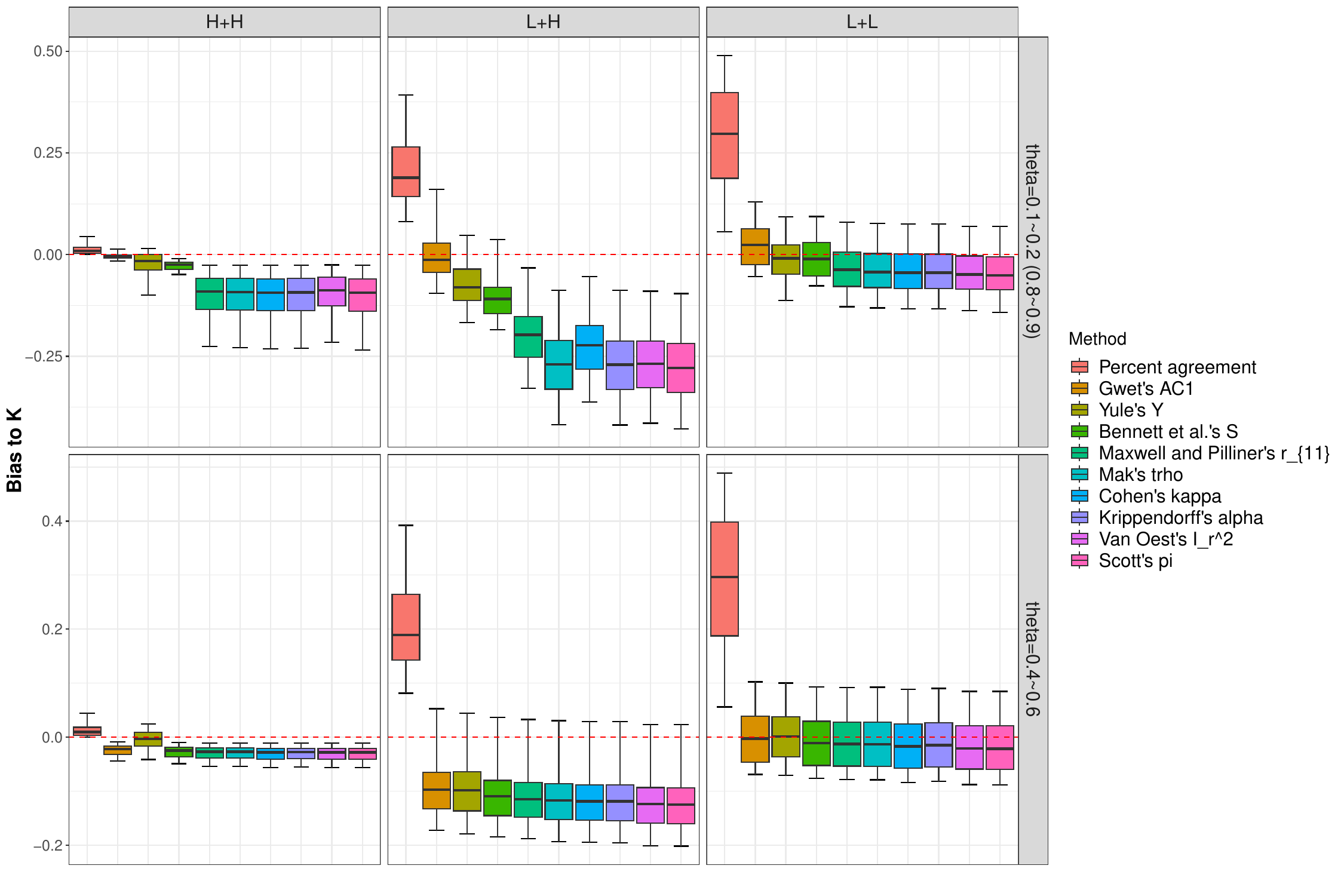}
\caption{Across-scenario overall bias to $K$ for $10$ interrater agreement measures at different joint rater-professionalism levels and typical outcome prevalence levels.} \label{fig: JPL_bias} 
\end{figure}

Figure \ref{fig: JPL_bias} shows the overall comparison by fixing rater characteristic combinations at certain hypothetical levels. We assume that those \textit{professional} raters tend to have both $p_i$ and $m_i$ low, $i=1,2$. This allows us to crudely define two levels of rater characteristics within our settings: raters with ``high'' professionalism ($p_i=0.1~\text{and}~0.3$ and $m_i=0.1~\text{and}~0.2$, denoted as ``H'') and raters with ``low'' professionalism ($p_i=0.7~\text{and}~0.9$ and $m_i=0.4~\text{and}~0.5$, denoted as ``L''). We present three levels of joint rater-professionalism combinations in Figure \ref{fig: JPL_bias}, from two ``high''-profession raters (H+H) to two ``low''-profession raters (L+L). When both raters' professionalism are acceptable, then $\text{AC}_1$ has the smallest median bias when the outcome prevalence is extreme, while $Y$ performs even better when the outcome prevalence is close to 0.5 and both raters are highly qualified. When both raters are highly-qualified and the outcome prevalence is extreme, the observed percent agreement shows a minor absolute bias to $K$ compared to that group of similarly-performed chance-corrected methods. Finally, when the raters are not well-qualified for the rating task, $\text{AC}_1$, $Y$, and $S$ are good choices that are slightly better than the method cluster from $r_{11}$ to $\pi$.

\begin{figure}[htbp!]
\centering
\includegraphics[width=1\textwidth]{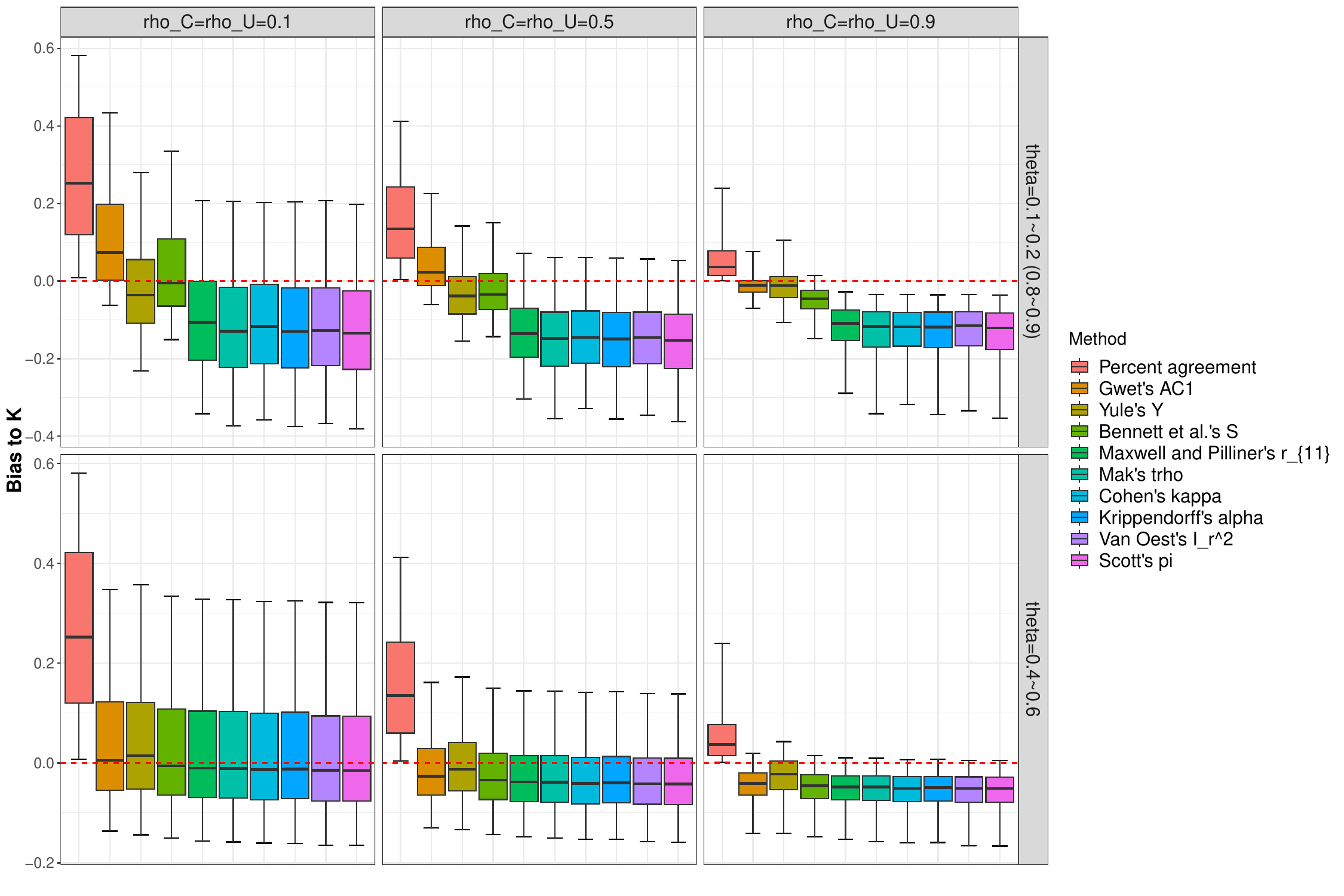}
\caption{Across-scenario overall bias to $K$ for $10$ interrater agreement measures at different rater correlation levels and typical outcome prevalence levels.} \label{fig: rho_bias} 
\end{figure}

Moreover, we aggregated the choices of $\rho_U$ and $\rho_C$ in Figure \ref{fig: rho_bias} to approximate low to high levels of overall rater behavioral correlation in a similar sense of parameter aggregation/truncation for $p_i$ and $m_i$. Specifically, we plot the across-scenario biases at $\rho_U=\rho_C=\{0.1, 0.5, 0.9\}$. $S$ possesses the smallest bias when the aggregated correlation is low. $\text{AC}_1$, $Y$, and $S$ generally outperform other methods regarding bias to $K$ at different levels of joint correlations and when the prevalence is extreme, while when the prevalence is close to 0.5, $Y$ is slightly better than all other IRA methods when the correlation parameters are in middle to high values. Along the increase of the joint latent correlations, the between-scenario variations for all the IRA methods seem to decrease.

\section{Discussion}\label{sec: 5}

We reviewed multiple chance-corrected IRA methods under the most common two-rater dichotomous-response case. To emphasize the mathematical connections among the considered IRA statistics, we introduced these measures in a special order and summarized them in a diagram and an equation table to facilitate the understanding and usage of these methods. To better mimic the 
real-world rating process, we proposed a generalized data-generating framework that introduces the underlying prevalence of the ``positive'' outcome of the rating subjects, and two rater-specific characteristics: the probability to feel \textit{uncertain} about the subject and the probability to make \textit{wrong} decision about the subject opposite against its ``true'' outcome. The concept of ``certainty'' and ``uncertainty'' is similar to the rater behavioral assumptions presented in \textcite{aickin1990maximum}, \textcite{gwet2008computing, gwet2014handbook}, and \textcite{van2019new}. However, different from their assumptions of making completely random votes when feeling \textit{uncertain}, we assume raters can still make rational judgement even when they feel \textit{uncertain}, which is controlled by the second decision step regarding making correct or wrong decisions. Another novelty of our generalized framework is that it allows the decision processes of the two raters to involve intrinsic correlations, which is induced by the facts that the raters work on the same set of subjects and they may receive similar training about the rating task.
Moreover, our generalized data-generating framework has the flexibility to reduce to simpler frameworks with different assumptions or beliefs about the two-rater dichotomous-response rating process. For example, when the two latent correlations are $\rho_U=\rho_C=0$ and the misclassification rates are $m_1=m_2=0.5$, then it reduces to the data-generating mechanism discussed in \textcite{gwet2008computing}. Furthermore, some researchers may question the assumption of certainty/uncertainty classification in the generating process of rating data. Our framework offers flexibility; it can be simplified without introducing the concepts of certainty and uncertainty by setting $p_1=p_2=1$ and eliminating the $\rho_U$ parameter. Some related discussions and simulation results can be found in Web Appendix E. Above all, our simulation study enables to comprehensively evaluate and compare a large list of IRA statistics under diverse and flexible conditions that might occur in practice. 

%Discussion about K
A rigorously defined estimand is the basis for making comparisons among different estimators for that ultimate target. However, in previous work, ``true'' IRA is seldom defined due to its conceptual vagueness and non-unified definitions of ``chance agreement'' in different literature \parencite{xu2014interrater,zhao2012reliability,aickin1990maximum}. Following the basic chance-correction IRA framework, we proposed a generalized ``true'' measure of chance-corrected IRA, named as $K$, to serve as the comparison benchmark in the simulation study. Compared with Gwet's ``true'' IRA definition, we had a more sensible classification of ``true agreement'' and ``chance agreement'' in the process of rating and agreement generation with potential decision correlations. We claimed that although ``chance agreement'' is born from the cases where at least one rater feels uncertain, a proportion of the agreement emerging under ``uncertainty'' are probabilistically deterministic, hence, should be counted as ``true agreement.'' In other words, the \textit{probabilistic certainty} that is induced by the underlying rater behavioral correlation should be accounted for under the case of ``uncertainty''. %Discuss the alternative definition of K and the potential issue
Regarding the approach of extracting ``true agreement'' from the at-least-one uncertainty case, an alternative ``chance agreement'' definition might be proposed as
\begin{align*}
    &p_e^*
    =P(\text{Agreement}~|~\text{At least one rater feel uncertain}~,~\rho_U=\rho_C=0)\\
    &=p_1p_2[m_1m_2+(1-m_1)(1-m_2)]+p_1(1-p_2)(1-m_1)+(1-p_1)p_2(1-m_2)+(1-p_1)(1-p_2)\times 0\\
    &=2p_1p_2m_1m_2+(1-m_1)p_1+(1-m_2)p_2-p_1p_2,
\end{align*}
and the ``true'' IRA is therefore defined as $K^*=(p_a-p_e^*)/(1-p_e^*)$. Readers familiar with the definition of Cohen's $\kappa$ might find the concept here to be quite analogous, where ``chance agreement'' is estimated conditional on the hypothesis of independent rating, or equivalent to the case of null agreement. However, this approach is somewhat counter-intuitive since it stands on the null hypothesis of no agreement, but in fact, there exists some deemed underlying correlation or some non-null agreement between the two raters. Instead, partitioning the ``total agreement'' reached under uncertainty and estimate the approximate proportion of the ``chance agreement'' subset (or equivalently, the ``true agreement'') is more appropriate for designing the estimand of ``true'' chance-corrected IRA measure.

%Discuss the findings from the overall assessment
As the ``true'' level of IRA can never be measured from real data and it is sometimes difficult to collect detailed rater or subject information in rating tasks, comparing IRA statistics cannot be accomplished with real datasets. By identifying a sensible truth of the target agreement estimate and conducting Monte Carlo simulations, both challenges above can be solved, and a comprehensive assessment can be completed by monitoring different IRA measures' performances across all possible levels of the key simulation parameters. By fixing those rater- or subject-related simulation factors at specific levels with practical interpretations, the simulation results can give useful guidance for selecting well-performed IRA measures under typical practical scenarios. When the information about the rater qualifications or the prevalence of ``positive'' outcome can hardly be obtained, we suggest using Gwet's $\text{AC}_1$, Yule's $Y$, or Bennett et al.'s $S$, as indicated in Figure \ref{fig: overall}. The agglomerative hierarchical clustering analysis also shows in an overall view that $\text{AC}_1$, $Y$, and $S$ are closer to our benchmark $K$ across various settings compared to the rest of IRA methods in our review. In many practical contexts, the prevalence level of the ``positive'' rating outcome could be estimated based on historical data, and it would also be the case that some information about rater characteristics could be garnered through professional credentials. With the knowledge on these two aspects, Figure \ref{fig: JPL_bias} could serve as a helpful resource to give more customized guidance for the optimal choice of IRA statistics. Specifically, when the two raters are both well-qualified for the rating task, the observed percent agreement can also be a good choice regarding the bias to the benchmark $K$, which makes the agreement estimation easy-to-implement. When the prevalence level is near $0.5$ or when the unobserved rater behavioral correlation is very small or very large, Yule's $Y$ or Bennett et al.'s $S$ provide better estimates regarding the bias to $K$. On the aspect of satisfactory bias performance as well as the stable performance of interval estimates, we generally recommend $\text{AC}_1$ and $S$ as the tools for IRA measurements in future practices. Our simulations also suggest that Maxwell and Pilliner's $r_{11}$, Mak's $\tilde{\rho}$, Cohen's $\kappa$, Krippendorff's $\alpha$, Van Oest's $\hat{I}_r^2$, and Scott's $\pi$ show similar bias and coverage performances in overall, and their overall clustering pattern is also validated through the cluster analysis. Importantly, these methods, including the famous Cohen's $\kappa$, tend to underestimate $K$ in most situations. Therefore, to develop new IRA methods targeting to this ``missing piece'' might be an essential direction of future work. In fact, our simulation results show that all the methods tend to cause notable bias against $K$ under certain scenarios, which suggests that, in our method comparison pool, there is no perfect method based on our $K$ definition, and future endeavors are needed in the methodological development of new agreement measures. 

There are some limitations in this comparative simulation study. In the current study, we focused on the rater characteristics and implicitly assumed that the objective difficulty in different subjects was constant. To make it more realistic in simulation settings, this limitation could be mitigated by including some subject covariates considering the task difficulty or intrinsic features of the rating items, but this would also introduce additional complexity in the data-generating framework. Moreover, we only included the reviewed variance estimators in the simulation study for comparison. Algorithmic methods, such as bootstrapping, can also be applied for every IRA statistic for interval estimation. Their performance, to be compared with the variance estimators in the literature, could be examined in future studies. Lastly, it is also possible to extend our simulation framework to the multiple-raters multiple-category outcome cases, which allows us to compare more popular measures like weighted kappa under the scenarios involving correlated rater and other realistic features. Nevertheless, insights have been provided in this study and consistent conclusions are anticipated, as the theoretical properties of these methods are identical regardless of the number of raters or number of categories.

\printbibliography

@book{agresti2003categorical,
  title={Categorical data analysis},
  author={Agresti, Alan},
  year={2003},
  publisher={John Wiley \& Sons}
}

@article{aickin1990maximum,
  title={Maximum likelihood estimation of agreement in the constant predictive probability model, and its relation to Cohen's kappa},
  author={Aickin, Mikel},
  journal={Biometrics},
  pages={293--302},
  year={1990},
  publisher={JSTOR}
}

@article{albert1993bayesian,
  title={Bayesian analysis of binary and polychotomous response data},
  author={Albert, James H and Chib, Siddhartha},
  journal={Journal of the American Statistical Association},
  volume={88},
  number={422},
  pages={669--679},
  year={1993},
  publisher={Taylor \& Francis}
}

@article{banerjee1999beyond,
  title={Beyond kappa: A review of interrater agreement measures},
  author={Banerjee, Mousumi and Capozzoli, Michelle and McSweeney, Laura and Sinha, Debajyoti},
  journal={Canadian Journal of Statistics},
  volume={27},
  number={1},
  pages={3--23},
  year={1999},
  publisher={Wiley Online Library}
}

@article{bartko1966intraclass,
  title={The intraclass correlation coefficient as a measure of reliability},
  author={Bartko, John J},
  journal={Psychological Reports},
  volume={19},
  number={1},
  pages={3--11},
  year={1966},
  publisher={SAGE Publications Sage CA: Los Angeles, CA}
}

@article{bennett1954communications,
  title={Communications through limited-response questioning},
  author={Bennett, Edward M and Alpert, Renee and Goldstein, AC},
  journal={Public Opinion Quarterly},
  volume={18},
  number={3},
  pages={303--308},
  year={1954},
  publisher={Oxford University Press}
}

@article{bexkens2018kappa,
  title={The kappa paradox},
  author={Bexkens, Rens and Claessen, Femke MAP and Kodde, Izaak F and Oh, Luke S and Eygendaal, Denise and den Bekerom, Michel PJ van},
  journal={Shoulder \& Elbow},
  volume={10},
  number={4},
  pages={308--308},
  year={2018},
  publisher={SAGE Publications Sage UK: London, England}
}

@article{blackman1993estimating,
  title={Estimating rater agreement in 2x2 tables: Correction for chance and intraclass correlation},
  author={Blackman, Nicole JM and Koval, John J},
  journal={Applied Psychological Measurement},
  volume={17},
  number={3},
  pages={211--223},
  year={1993},
  publisher={Sage Publications Sage CA: Thousand Oaks, CA}
}

@article{bloch19892,
  title={2x2 kappa coefficients: measures of agreement or association},
  author={Bloch, Daniel A and Kraemer, Helena Chmura},
  journal={Biometrics},
  pages={269--287},
  year={1989},
  publisher={JSTOR}
}

@article{bonett2007statistical,
  title={Statistical inference for generalized Yule coefficients in 2x2 contingency tables},
  author={Bonett, Douglas G and Price, Robert M},
  journal={Sociological Methods and Research},
  volume={35},
  number={3},
  pages={429--446},
  year={2007},
  publisher={Sage Publications Sage CA: Thousand Oaks, CA}
}

@article{brennan1981coefficient,
  title={Coefficient kappa: Some uses, misuses, and alternatives},
  author={Brennan, Robert L and Prediger, Dale J},
  journal={Educational and Psychological Measurement},
  volume={41},
  number={3},
  pages={687--699},
  year={1981},
  publisher={Sage Publications Sage CA: Thousand Oaks, CA}
}

@article{byrt1993bias,
  title={Bias, prevalence and kappa},
  author={Byrt, Ted and Bishop, Janet and Carlin, John B},
  journal={Journal of Clinical Epidemiology},
  volume={46},
  number={5},
  pages={423--429},
  year={1993},
  publisher={Elsevier}
}

@article{Choudhary2005,
author="Choudhary, Pankaj K.
and Nagaraja, H. N.",
editor="Balakrishnan, N.
and Nagaraja, H. N.
and Kannan, N.",
title="Measuring Agreement in Method Comparison Studies --- A Review",
journal={Statistics for Industry and Technologys},
year={2005},
publisher={Birkh{\"a}user Boston},
pages={215--244}
}

@article{cicchetti1977comparison,
  title={Comparison of the null distributions of weighted kappa and the C ordinal statistic},
  author={Cicchetti, Dominic V and Fleiss, Joseph L},
  journal={Applied Psychological Measurement},
  volume={1},
  number={2},
  pages={195--201},
  year={1977},
  publisher={Sage Publications Sage CA: Thousand Oaks, CA}
}

@article{cicchetti1990high,
  title={High agreement but low kappa: II. Resolving the paradoxes},
  author={Cicchetti, Domenic V and Feinstein, Alvan R},
  journal={Journal of Clinical Epidemiology},
  volume={43},
  number={6},
  pages={551--558},
  year={1990},
  publisher={Elsevier}
}

@article{cohen1960coefficient,
  title={A coefficient of agreement for nominal scales},
  author={Cohen, Jacob},
  journal={Educational and Psychological Measurement},
  volume={20},
  number={1},
  pages={37--46},
  year={1960},
  publisher={Sage Publications Sage CA: Thousand Oaks, CA}
}

@article{feinstein1990high,
  title={High agreement but low kappa: I. The problems of two paradoxes},
  author={Feinstein, Alvan R and Cicchetti, Domenic V},
  journal={Journal of Clinical Epidemiology},
  volume={43},
  number={6},
  pages={543--549},
  year={1990},
  publisher={Elsevier}
}

@article{feng2013factors,
  title={Factors affecting intercoder reliability: A Monte Carlo experiment},
  author={Feng, Guangchao Charles},
  journal={Quality \& Quantity},
  volume={47},
  number={5},
  pages={2959--2982},
  year={2013},
  publisher={Springer}
}

@article{feng2013underlying,
  title={Underlying determinants driving agreement among coders},
  author={Feng, Guangchao Charles},
  journal={Quality \& Quantity},
  volume={47},
  number={5},
  pages={2983--2997},
  year={2013},
  publisher={Springer}
}

@article{finn1970note,
  title={A note on estimating the reliability of categorical data},
  author={Finn, Robert H},
  journal={Educational and Psychological Measurement},
  volume={30},
  number={1},
  pages={71--76},
  year={1970},
  publisher={Sage Publications Sage CA: Thousand Oaks, CA}
}

@article{fleiss1969large,
  title={Large sample standard errors of kappa and weighted kappa.},
  author={Fleiss, Joseph L and Cohen, Jacob and Everitt, Brian S},
  journal={Psychological Bulletin},
  volume={72},
  number={5},
  pages={323},
  year={1969},
  publisher={American Psychological Association}
}

@article{fleiss1971measuring,
  title={Measuring nominal scale agreement among many raters.},
  author={Fleiss, Joseph L},
  journal={Psychological Bulletin},
  volume={76},
  number={5},
  pages={378},
  year={1971},
  publisher={American Psychological Association}
}

@article{fleiss1975measuring,
  title={Measuring agreement between two judges on the presence or absence of a trait},
  author={Fleiss, Joseph L},
  journal={Biometrics},
  pages={651--659},
  year={1975},
  publisher={JSTOR}
}

@article{fleiss1978inference,
  title={Inference about weighted kappa in the non-null case},
  author={Fleiss, Joseph L and Cicchetti, Domenic V},
  journal={Applied Psychological Measurement},
  volume={2},
  number={1},
  pages={113--117},
  year={1978},
  publisher={Sage Publications Sage CA: Thousand Oaks, CA}
}

@article{gisev2013interrater,
  title={Interrater agreement and interrater reliability: key concepts, approaches, and applications},
  author={Gisev, Natasa and Bell, J Simon and Chen, Timothy F},
  journal={Research in Social and Administrative Pharmacy},
  volume={9},
  number={3},
  pages={330--338},
  year={2013},
  publisher={Elsevier}
}

@article{grant2017evaluation,
  title={An evaluation of interrater reliability measures on binary tasks using d-prime},
  author={Grant, Malcolm J and Button, Cathryn M and Snook, Brent},
  journal={Applied Psychological Measurement},
  volume={41},
  number={4},
  pages={264--276},
  year={2017},
  publisher={Sage Publications Sage CA: Los Angeles, CA}
}

@article{grove1981reliability,
  title={Reliability studies of psychiatric diagnosis: Theory and practice},
  author={Grove, William M and Andreasen, Nancy C and McDonald-Scott, Patricia and Keller, Martin B and Shapiro, Robert W},
  journal={Archives of General Psychiatry},
  volume={38},
  number={4},
  pages={408--413},
  year={1981},
  publisher={American Medical Association}
}

@article{gwet2008computing,
  title={Computing inter-rater reliability and its variance in the presence of high agreement},
  author={Gwet, Kilem Li},
  journal={British Journal of Mathematical and Statistical Psychology},
  volume={61},
  number={1},
  pages={29--48},
  year={2008},
  publisher={Wiley Online Library}
}

@book{gwet2014handbook,
  title={Handbook of inter-rater reliability: The definitive guide to measuring the extent of agreement among raters},
  author={Gwet, Kilem L},
  year={2021},
  publisher={Advanced Analytics, LLC}
}

@article{hayes2007answering,
  title={Answering the call for a standard reliability measure for coding data},
  author={Hayes, Andrew F and Krippendorff, Klaus},
  journal={Communication Methods and Measures},
  volume={1},
  number={1},
  pages={77--89},
  year={2007},
  publisher={Taylor \& Francis}
}

@article{hoehler2000bias,
  title={Bias and prevalence effects on kappa viewed in terms of sensitivity and specificity},
  author={Hoehler, Fred K},
  journal={Journal of Clinical Epidemiology},
  volume={53},
  number={5},
  pages={499--503},
  year={2000},
  publisher={Elsevier}
}

@article{holley1964note,
  title={A note on the G index of agreement},
  author={Holley, Jasper Wilson and Guilford, Joy Paul},
  journal={Educational and Psychological Measurement},
  volume={24},
  number={4},
  pages={749--753},
  year={1964},
  publisher={Sage Publications Sage CA: Thousand Oaks, CA}
}

@article{janson1979generalizations,
  title={On generalizations of the G index and the phi coefficient to nominal scales},
  author={Janson, Svante and Vegelius, Jan},
  journal={Multivariate Behavioral Research},
  volume={14},
  number={2},
  pages={255--269},
  year={1979},
  publisher={Taylor \& Francis}
}

@book{kazdin2021research,
  title={Research design in clinical psychology},
  author={Kazdin, Alan E},
  year={2021},
  publisher={Cambridge University Press}
}

@article{krippendorff1970bivariate,
  title={Bivariate agreement coefficients for reliability of data},
  author={Krippendorff, Klaus},
  journal={Sociological Methodology},
  volume={2},
  pages={139--150},
  year={1970},
  publisher={JSTOR}
}

@article{landis1975review,
  title={A review of statistical methods in the analysis of data arising from observer reliability studies (Part I)},
  author={Landis, J Richard and Koch, Gary G},
  journal={Statistica Neerlandica},
  volume={29},
  number={3},
  pages={101--123},
  year={1975},
  publisher={Wiley Online Library}
}

@article{m2020Interobserver,
  title={Interobserver agreement issues in radiology},
  author={Benchoufi, Mehdi and Matzner-Lober, Eric and Molinari, Nicolas and Jannot, Anne Sophie and Soyer, P},
  journal={Diagnostic and Interventional Imaging},
  volume={101},
  number={10},
  pages={639--641},
  year={2020},
  publisher={Elsevier}
}

@article{mak1988analysing,
  title={Analysing intraclass correlation for dichotomous variables},
  author={Mak, Tak K},
  journal={Journal of the Royal Statistical Society: Series C (Applied Statistics)},
  volume={37},
  number={3},
  pages={344--352},
  year={1988},
  publisher={Wiley Online Library}
}

@article{marshall1985family,
  title={A family of bivariate distributions generated by the bivariate Bernoulli distribution},
  author={Marshall, Albert W and Olkin, Ingram},
  journal={Journal of the American Statistical Association},
  volume={80},
  number={390},
  pages={332--338},
  year={1985},
  publisher={Taylor \& Francis}
}

@article{maxwell1968deriving,
  title={Deriving coefficients of reliability and agreement for ratings},
  author={Maxwell, AE and Pilliner, AEG},
  journal={British Journal of Mathematical and Statistical Psychology},
  volume={21},
  number={1},
  pages={105--116},
  year={1968},
  publisher={Wiley Online Library}
}

@article{maxwell1977coefficients,
  title={Coefficients of agreement between observers and their interpretation},
  author={Maxwell, Ann E},
  journal={The British Journal of Psychiatry},
  volume={130},
  number={1},
  pages={79--83},
  year={1977},
  publisher={Cambridge University Press}
}

@article{perreault1989reliability,
  title={Reliability of nominal data based on qualitative judgments},
  author={Perreault Jr, William D and Leigh, Laurence E},
  journal={Journal of Marketing Research},
  volume={26},
  number={2},
  pages={135--148},
  year={1989},
  publisher={SAGE Publications Sage CA: Los Angeles, CA}
}

@article{potter1999rethinking,
  title={Rethinking validity and reliability in content analysis},
  author={Potter, W James and Levine-Donnerstein, Deborah},
  year={1999},
  publisher={Taylor \& Francis}
}

@article{scott1955reliability,
  title={Reliability of content analysis: The case of nominal scale coding},
  author={Scott, William A},
  journal={Public Opinion Quarterly},
  pages={321--325},
  year={1955},
  publisher={JSTOR}
}

@article{spitznagel1985proposed,
  title={A proposed solution to the base rate problem in the kappa statistic},
  author={Spitznagel, Edward L and Helzer, John E},
  journal={Archives of General Psychiatry},
  volume={42},
  number={7},
  pages={725--728},
  year={1985},
  publisher={American Medical Association}
}

@Article{sym14020262,
AUTHOR = {Konstantinidis, Menelaos and Le, Lisa. W. and Gao, Xin},
TITLE = {An Empirical Comparative Assessment of Inter-Rater Agreement of Binary Outcomes and Multiple Raters},
JOURNAL = {Symmetry},
VOLUME = {14},
YEAR = {2022},
NUMBER = {2},
ARTICLE-NUMBER = {262},
ISSN = {2073-8994}
}

@article{vach2005dependence,
  title={The dependence of Cohen's kappa on the prevalence does not matter},
  author={Vach, Werner},
  journal={Journal of Clinical Epidemiology},
  volume={58},
  number={7},
  pages={655--661},
  year={2005},
  publisher={Elsevier}
}

@article{van2019new,
  title={A new coefficient of interrater agreement: The challenge of highly unequal category proportions},
  author={Van Oest, Rutger},
  journal={Psychological Methods},
  volume={24},
  number={4},
  pages={439},
  year={2019},
  publisher={American Psychological Association}
}

@article{walter2001hoehler,
  title={Hoehler's adjusted kappa is equivalent to Yule's Y},
  author={Walter, SD},
  journal={Journal of Clinical Epidemiology},
  volume={54},
  number={10},
  pages={1072},
  year={2001},
  publisher={Elsevier}
}

@article{xu2014interrater,
  title={Interrater agreement statistics with skewed data: Evaluation of alternatives to Cohen’s kappa},
  author={Xu, Shu and Lorber, Michael F},
  journal={Journal of Consulting and Clinical Psychology},
  volume={82},
  number={6},
  pages={1219},
  year={2014},
  publisher={American Psychological Association}
}

@article{yule1912methods,
  title={On the methods of measuring association between two attributes},
  author={Yule, G Udny},
  journal={Journal of the Royal Statistical Society},
  volume={75},
  number={6},
  pages={579--652},
  year={1912},
  publisher={JSTOR}
}

@article{zhao2012reliability,
  title={A Reliability Index (ai) that assumes honest coders and variable randomness},
  author={Zhao, Xinshu},
  year={2012}
}

@article{zhao2013assumptions,
  title={Assumptions behind intercoder reliability indices},
  author={Zhao, Xinshu and Liu, Jun S and Deng, Ke},
  journal={Annals of the International Communication Association},
  volume={36},
  number={1},
  pages={419--480},
  year={2013},
  publisher={Taylor \& Francis}
}

@article{zwick1988another,
  title={Another look at interrater agreement.},
  author={Zwick, Rebecca},
  journal={Psychological Bulletin},
  volume={103},
  number={3},
  pages={374},
  year={1988},
  publisher={American Psychological Association}
}

%\section{Conclusion}

%This study reviewed a series of popularly used chance-corrected IRA measures and developed a novel simulation scheme to evaluate a large list of IRA statistics under the common two-rater dichotomous-response case. By generalizing some assumptions in previous simulation studies and involving more comprehensive factors about the raters and rating task, a new data-generating framework was proposed, and a more reasonable ``true'' chance-corrected IRA benchmark was constructed using the key simulation parameters. Our extensive simulation study provides useful implications on the behavioral patterns of IRA measures and also offers practical suggestions about the optimal choices of IRA under different practical scenarios. We believe this study can not only help the investigators in practice to proceed more scientifically but also leave insights for developing improved IRA statistics in future methodological studies. 

%\bibliographystyle{plain}
%\bibliography{sample}

%\begin{spacing}{1.5}

%\end{spacing}

\clearpage

\section*{Web Appendix A: Interrater agreement statistics with intraclass correlation coefficient interpretations}

The Analysis of Variance (ANOVA) model is a standard way to assess the level of rater variability when the outcome is continuous. A basic linear mixed effects model with ``subject effect'' but no ``rater effect'' (or understood as including the ``rater effect'' in the residual error term) can be written as
\begin{equation}\label{eq: lme1}
    y_{ij}=\mu+s_i+e_{ij},
\end{equation}
where $y_{ij}$ is a continuous variable but with data inputs $\{0,1\}$, $i=1,2,...,N$ represents the index of rating subjects, and $j=1,2,...,d$ represents the indicator of rater. Assume the $N$ subjects are a random sample from a population, $s_i\sim N(0,\sigma_s^2)$, $e_{ij}\sim N(0, \sigma_e^2)$, and $s_i$ and $e_{ij}$ are independent. Under this model, the intraclass correlation coefficient (ICC) or the correlation of raters on the same subject is a measure of rater agreement on the same sets of subjects, which is denoted by
\begin{equation*}
    q_1=\frac{\text{Cov}(y_{i1},y_{i2})}{\sqrt{\text{Var}(y_{i1})}\sqrt{\text{Var}(y_{i2})}}=\frac{\sigma_s^2}{\sigma_s^2+\sigma_e^2}.
\end{equation*}
Regarding the model, the ANOVA table can be constructed as
\begin{table}[!th]
\caption{ANOVA table for model \eqref{eq: lme1}}\label{tb: ANOVA1}
\begin{center}
\begin{tabular}{ |c|c|c|c| }
\hline
Source & df & SS & MS \\ 
\hline
\multirow{3}{*}{Subject (b/w subject)} 
& & & \\
& $N-1$ & $SS_s=\sum_{i,j}(\bar{y}_{i.}-\bar{y})^2$ & $MS_s=\frac{1}{N-1}SS_s$\\
& & & \\
\multirow{3}{*}{Error (w/i subject)} 
& & & \\
& $N(d-1)$ & $SS_e=\sum_{i,j}(y_{ij}-\bar{y}_{i.})^2$ & $MS_e=\frac{1}{N(d-1)}SS_e$\\
& & & \\
\hline
\multirow{3}{*}{Total}
 & & & \\
 & $Nd-1$ & $SS_t=\sum_{i,j}(y_{ij}-\bar{y})^2$ & \\
 & & &\\
\hline
\end{tabular}
\end{center}
\end{table}

Then, the unbiased and consistent estimates for the variance components are given by $\hat{\sigma}_s^2=(MS_s-MS_e)/d$, $\hat{\sigma}_e^2=MS_e$, and a consistent estimate of $q_1$ is given by
\begin{equation*}
    \hat{q}_1=\frac{\hat{\sigma}_s^2}{\hat{\sigma}_s^2+\hat{\sigma}_e^2}=\frac{MS_s-MS_e}{MS_s+(d-1)MS_e}.
\end{equation*}

By setting the number of raters $d=2$, the ANOVA table can be expressed with the estimated proportions in the common $2\times 2$ contingency table. Also, it follows $y_{ij}^2=y_{ij}$. It can be shown that
\begin{align*}
    &\sum_{i,j}y_{ij} = 2N \bar{y} = \sum_i (y_{i1}+y_{i2}) = 2n_{11}+n_{10}+n_{01}=N(2\hat{p}_{11}+\hat{p}_{10}+\hat{p}_{01}),\\
    &\sum_i y_{i1}y_{i2} = a=N\hat{p}_{11},\\
    &\sum_i \bar{y}_{i.}^2 =\sum_i \left[\frac{1}{2}(y_{i1}+y_{i2})\right]^2=\frac{1}{4}\sum_i(y_{i1}+y_{i2}+2y_{i1}y_{i2})=\frac{N}{4}(4\hat{p}_{11}+\hat{p}_{10}+\hat{p}_{01}),\\
    &SS_s =2\sum_i \bar{y}_{i.}^2-2N\bar{y}^2=\frac{N}{2}(4\hat{p}_{11}+\hat{p}_{10}+\hat{p}_{01})-\frac{N}{2}(2\hat{p}_{11}+\hat{p}_{10}+\hat{p}_{01})^2=\frac{N}{2}[(\hat{p}_{11}+\hat{p}_{00})-(\hat{p}_{11}-\hat{p}_{00})^2],\\
    &SS_e =\sum_{i,j}y_{ij}-2\sum_i \bar{y}_{i.}^2=N(2\hat{p}_{11}+\hat{p}_{10}+\hat{p}_{01})-\frac{N}{2}(4\hat{p}_{11}+\hat{p}_{10}+\hat{p}_{01})=\frac{N}{2}(\hat{p}_{10}+\hat{p}_{01}),
\end{align*}
then,
\begin{align*}
    \hat{q}_1
    &=\frac{MS_s-MS_e}{MS_s+MS_e}\\
    &=\frac{\frac{N}{N-1}[(\hat{p}_{11}+\hat{p}_{00})-(\hat{p}_{11}-\hat{p}_{00})^2]-(\hat{p}_{10}+\hat{p}_{01})}{\frac{N}{N-1}[(\hat{p}_{11}+\hat{p}_{00})-(\hat{p}_{11}-\hat{p}_{00})^2]+(\hat{p}_{10}+\hat{p}_{01})},\\
    &=\frac{N[4\hat{p}_{11}\hat{p}_{00}+(\hat{p}_{11}+\hat{p}_{00})(\hat{p}_{10}+\hat{p}_{01})]-(N-1)(\hat{p}_{10}+\hat{p}_{01})}{N[4\hat{p}_{11}\hat{p}_{00}+(\hat{p}_{11}+\hat{p}_{00})(\hat{p}_{10}+\hat{p}_{01})]+(N-1)(\hat{p}_{10}+\hat{p}_{01})},\\
    &=\frac{N[4\hat{p}_{11}\hat{p}_{00}+(\hat{p}_{11}+\hat{p}_{00}-1)(\hat{p}_{10}+\hat{p}_{01})]+(\hat{p}_{10}+\hat{p}_{01})}{N[4\hat{p}_{11}\hat{p}_{00}+2(\hat{p}_{11}+\hat{p}_{00})(\hat{p}_{10}+\hat{p}_{01})]+N(\hat{p}_{10}+\hat{p}_{01})^2-(\hat{p}_{10}+\hat{p}_{01})},\\
    &=\frac{4(\hat{p}_{11}\hat{p}_{00}-\hat{p}_{10}\hat{p}_{01})-(\hat{p}_{10}-\hat{p}_{01})^2+\frac{1}{N}(\hat{p}_{10}+\hat{p}_{01})}{(2\hat{p}_{11}+\hat{p}_{10}+\hat{p}_{01})(2\hat{p}_{00}+\hat{p}_{10}+\hat{p}_{01})-\frac{1}{N}(\hat{p}_{10}+\hat{p}_{01})}=\tilde{\rho},
\end{align*}
which shows an exact equality to Mak's $\tilde{\rho}$ under the $2\times 2$ case. And, it follows that $\hat{q}_1$ is asymptotically equal to Scott's $\pi$, see
\begin{align*}
    \hat{q}_1\rightarrow \frac{4(\hat{p}_{11}\hat{p}_{00}-\hat{p}_{10}\hat{p}_{01})-(\hat{p}_{10}-\hat{p}_{01})^2}{(2\hat{p}_{11}+\hat{p}_{10}+\hat{p}_{01})(2\hat{p}_{00}+\hat{p}_{00}+\hat{p}_{01})}=\pi,~~\text{as}~N\rightarrow \infty.
\end{align*}

Next, a model additionally considering the ``rater effect'' can be written as
\begin{equation}\label{eq: lme2}
    y_{ij}=\mu+s_i+d_j+e_{ij},
\end{equation}
with regular assumptions similarly set for model \eqref{eq: lme1}, plus $d_j\sim N(0,\sigma_d^2)$. Note that the $d$ raters are assumed to be a random sample from a large population of potential raters of interest. Under this model, the within-subject intraclass correlation coefficient is given as
\begin{equation*}
    q_2=\frac{\sigma_s^2}{\sigma_s^2+\sigma_d^2+\sigma_e^2}.
\end{equation*}
Regarding this model, the ANOVA table can be constructed as
\begin{table}[!th]
\caption{ANOVA table for model \eqref{eq: lme2} and \eqref{eq: lme3}}\label{tb: ANOVA2}
\begin{center}
\begin{tabular}{ |c|c|c|c| }
\hline
Source & df & SS & MS \\ 
\hline
\multirow{3}{*}{rater (b/w rater)} 
& & & \\
& $d-1$ & $SS_d=\sum_{i,j}(\bar{y}_{.j}-\bar{y})^2$ & $MS_d=\frac{1}{d-1}SS_d$\\
& & & \\
\multirow{3}{*}{Subject (b/w subject)} 
& & & \\
& $N-1$ & $SS_s=\sum_{i,j}(\bar{y}_{i.}-\bar{y})^2$ & $MS_s=\frac{1}{N-1}SS_s$\\
& & & \\
\multirow{3}{*}{Error} 
& & & \\
& $(N-1)(d-1)$ & $SS_e=\sum_{i,j}(y_{ij}-\bar{y}_{i.}-\bar{y}_{.j}+\bar{y})^2$ & $MS_e=\frac{1}{(N-1)(d-1)}SS_e$\\
& & & \\
\hline
\multirow{3}{*}{Total}
 & & & \\
 & $nd-1$ & $SS_t=\sum_{i,j}(y_{ij}-\bar{y})^2$ & \\
 & & &\\
\hline
\end{tabular}
\end{center}
\end{table}

The unbiased and consistent estimates for the variance components are given as 
\begin{align*}
    \hat{\sigma}_d^2=(MS_d-MS_e)/N,~~\hat{\sigma}_s^2=(MS_s-MS_e)/d,~~\hat{\sigma}_e^2=MS_e,
\end{align*}
and a consistent estimate of $q_2$ is given by
\begin{equation*}
    \hat{q}_2=\frac{\hat{\sigma}_s^2}{\hat{\sigma}_s^2+\hat{\sigma}_d^2+\hat{\sigma}_e^2}=\frac{N(MS_s-MS_e)}{d(MS_d-MS_e)+N(MS_s-MS_e)+NdMS_e}.
\end{equation*}

By treating $y_{ij}$ as $\{0,1\}$-value variable and setting $d=2$, the ANOVA table can be expressed with proportions in $2\times 2$ table. It can be shown that
\begin{align*}
    \sum_j\bar{y}_{.j}^2&=\bar{y}_{.1}^2+\bar{y}_{.2}^2=\frac{1}{N^2}[(\sum_iy_{i1})^2+(\sum_iy_{i2})^2]=\frac{(n_{11}+n_{10})^2+(n_{11}+n_{01})^2}{N^2}=(\hat{p}_{11}+\hat{p}_{10})^2+(\hat{p}_{11}+\hat{p}_{01})^2,\\
    SS_d &=N\sum_j\bar{y}_{.j}^2-2N\bar{y}^2=N(\hat{p}_{11}+\hat{p}_{10})^2+N(\hat{p}_{11}+\hat{p}_{01})^2-\frac{N}{2}(2\hat{p}_{11}+\hat{p}_{10}+\hat{p}_{01})^2=\frac{N}{2}(\hat{p}_{10}-\hat{p}_{01})^2,\\
    SS_s &=2\sum_i \bar{y}_{i.}^2-2N\bar{y}^2=\frac{N}{2}[(\hat{p}_{11}+\hat{p}_{00})-(\hat{p}_{11}-\hat{p}_{00})^2],\\
    SS_e &=\sum_{i,j}y_{ij}-2\sum_i \bar{y}_{i.}^2=\frac{N}{2}[(\hat{p}_{10}+\hat{p}_{01})-(\hat{p}_{10}-\hat{p}_{01})^2],
\end{align*}
then,
\begin{align*}
    \hat{q}_2
    &=\frac{N(MS_s-MS_e)}{2MS_d+nMS_s+(N-2)MS_e}\\
    &=\frac{\frac{N}{N-1}[(\hat{p}_{11}+\hat{p}_{00})-(\hat{p}_{11}-\hat{p}_{00})^2]-\frac{N}{N-1}[(\hat{p}_{10}+\hat{p}_{01})-(\hat{p}_{10}-\hat{p}_{01})^2]}{2(\hat{p}_{10}-\hat{p}_{01})^2+\frac{N}{(N-1)}[(\hat{p}_{11}+\hat{p}_{00})-(\hat{p}_{11}-\hat{p}_{00})^2]+\frac{N-2}{N-1}[(\hat{p}_{10}+\hat{p}_{01})-(\hat{p}_{10}-\hat{p}_{01})^2]}\\
    &=\frac{\frac{N}{N-1}[4(\hat{p}_{11}\hat{p}_{00}-\hat{p}_{10}\hat{p}_{01})]}{2(\hat{p}_{10}-\hat{p}_{01})^2+\frac{N}{(N-1)}[2(\hat{p}_{11}+\hat{p}_{00})(\hat{p}_{10}+\hat{p}_{01})+4\hat{p}_{11}\hat{p}_{00}+4\hat{p}_{10}\hat{p}_{01}]-\frac{2}{N-1}[(\hat{p}_{10}+\hat{p}_{01})-(\hat{p}_{10}-\hat{p}_{01})^2]}\\
    &=\frac{2(\hat{p}_{11}\hat{p}_{00}-\hat{p}_{10}\hat{p}_{01})}{\frac{N-1}{N}(\hat{p}_{10}-\hat{p}_{01})^2+(\hat{p}_{11}+\hat{p}_{00})(\hat{p}_{10}+\hat{p}_{01})+2\hat{p}_{11}\hat{p}_{00}+2\hat{p}_{10}\hat{p}_{01}-\frac{1}{N}[(\hat{p}_{10}+\hat{p}_{01})-(\hat{p}_{10}-\hat{p}_{01})^2]}.
\end{align*}
This complicated measure do not have a formal name for IRA statistics at present. But, it 's easy to see that $\hat{q}_2$ is asymptotically equal to Cohen's $\kappa$,
\begin{align*}
    \hat{q}_2
    &\rightarrow \frac{2(\hat{p}_{11}\hat{p}_{00}-\hat{p}_{10}\hat{p}_{01})}{(\hat{p}_{10}-\hat{p}_{01})^2+(\hat{p}_{11}+\hat{p}_{00})(\hat{p}_{10}+\hat{p}_{01})+2\hat{p}_{11}\hat{p}_{00}+2\hat{p}_{10}\hat{p}_{01}}\\
    &=\frac{2(\hat{p}_{11}\hat{p}_{00}-\hat{p}_{10}\hat{p}_{01})}{(\hat{p}_{11}+\hat{p}_{10})(\hat{p}_{10}+\hat{p}_{00})+(\hat{p}_{11}+\hat{p}_{01})(\hat{p}_{01}+\hat{p}_{00})}=\kappa,~~\text{as}~N\rightarrow \infty.
\end{align*}

Based on model \eqref{eq: lme2}, if further assuming that the $d$ raters are from a fixed set, then the model can become
\begin{equation}\label{eq: lme3}
    y_{ij}=\mu+s_i+\delta_j+e_{ij},~~\sum^d_{j=1} \delta_j=1.
\end{equation}
The target ICC measure will change to
\begin{align*}
    q_3=\frac{\sigma_s^2}{\sigma_s^2+\sigma_e^2},
\end{align*}
with the unbiased and consistent variance component estimates unchanged as above under the setting of model \eqref{eq: lme2}.

Using the same ANOVA table derived for model \eqref{eq: lme2} (Table \ref{tb: ANOVA2}) and under the case of $d=2$, we can get the exact equivalence between Maxwell and Pilliner's $r_{11}$ and the consistent estimator of $q_3$:
\begin{align*}
    \hat{q}_3
    &=\frac{MS_s-MS_e}{MS_s+MS_e}\\
    &=\frac{\frac{N}{2(N-1)}[(\hat{p}_{11}+\hat{p}_{00})-(\hat{p}_{11}-\hat{p}_{00})^2]-\frac{N}{2(N-1)}[(\hat{p}_{10}+\hat{p}_{01})-(\hat{p}_{10}-\hat{p}_{01})^2]}{\frac{N}{2(N-1)}[(\hat{p}_{11}+\hat{p}_{00})-(\hat{p}_{11}-\hat{p}_{00})^2]+\frac{N}{2(N-1)}[(\hat{p}_{10}+\hat{p}_{01})-(\hat{p}_{10}-\hat{p}_{01})^2]}\\
    &=\frac{2(\hat{p}_{11}\hat{p}_{00}-\hat{p}_{10}\hat{p}_{01})}{(\hat{p}_{11}+\hat{p}_{00})(\hat{p}_{10}+\hat{p}_{01})+2\hat{p}_{11}\hat{p}_{00}+2\hat{p}_{10}\hat{p}_{01}}\\
    &=\frac{2(\hat{p}_{11}\hat{p}_{00}-\hat{p}_{10}\hat{p}_{01})}{(\hat{p}_{11}+\hat{p}_{10})(\hat{p}_{01}+\hat{p}_{00})+(\hat{p}_{11}+\hat{p}_{01})(\hat{p}_{10}+\hat{p}_{00})}=r_{11}
\end{align*}

Finally, a table summarizing the model assumptions and the relationships among the IRA statistics with ICC interpretations is provided as follows.
\begin{table}[htbp]
\begin{center}
\caption{Interrater agreement (IRA) statistics with intraclass correlation coefficient interpretations and their correspondences with different mixed effects/ANOVA models}\label{tb: IRA-ICC}
\scalebox{0.9}{
\begin{tabular}{ |c|c|c|c| }
\hline
Model assumption & IRA statistics & Calculation formula \\
\hline
\multirow{5}{*}{No rater-specific effect$^a$} 
& & \\
& Mak's $\tilde{\rho}$ & $\frac{4(\hat{p}_{11}\hat{p}_{00}-\hat{p}_{10}\hat{p}_{01})-(\hat{p}_{10}-\hat{p}_{01})^2+\frac{1}{N}(\hat{p}_{10}+\hat{p}_{01})}{(2\hat{p}_{11}+\hat{p}_{10}+\hat{p}_{01})(2\hat{p}_{00}+\hat{p}_{10}+\hat{p}_{01})-\frac{1}{N}(\hat{p}_{10}+\hat{p}_{01})}$ \\
& & \\
& Scott's $\pi$ & $\frac{4(\hat{p}_{11}\hat{p}_{00}-\hat{p}_{10}\hat{p}_{01})-(\hat{p}_{10}-\hat{p}_{01})^2}{(2\hat{p}_{11}+\hat{p}_{10}+\hat{p}_{01})(2\hat{p}_{00}+\hat{p}_{00}+\hat{p}_{01})}$ \\
& & \\
\hline
\multirow{5}{*}{Random rater effects$^b$} 
& & \\
& (No formal name) & $\frac{2(\hat{p}_{11}\hat{p}_{00}-\hat{p}_{10}\hat{p}_{01})}{\frac{N-1}{N}(\hat{p}_{10}-\hat{p}_{01})^2+(\hat{p}_{11}+\hat{p}_{00})(\hat{p}_{10}+\hat{p}_{01})+2\hat{p}_{11}\hat{p}_{00}+2\hat{p}_{10}\hat{p}_{01}-\frac{1}{N}[(\hat{p}_{10}+\hat{p}_{01})-(\hat{p}_{10}-\hat{p}_{01})^2]}$ \\
& & \\
& Cohen's $\kappa$ & $\frac{2(\hat{p}_{11}\hat{p}_{00}-\hat{p}_{10}\hat{p}_{01})}{(\hat{p}_{11}+\hat{p}_{10})(\hat{p}_{10}+\hat{p}_{00})+(\hat{p}_{11}+\hat{p}_{01})(\hat{p}_{01}+\hat{p}_{00})}$ \\
& & \\
\hline
\multirow{3}{*}{Raters from a fixed set$^c$} 
& & \\
& Maxwell and Pilliner's $r_{11}$ & $\frac{2(\hat{p}_{11}\hat{p}_{00}-\hat{p}_{10}\hat{p}_{01})}{(\hat{p}_{11}+\hat{p}_{10})(\hat{p}_{01}+\hat{p}_{00})+(\hat{p}_{11}+\hat{p}_{01})(\hat{p}_{10}+\hat{p}_{00})}$ \\
& &\\
\hline
\end{tabular}}
\end{center}
\footnotesize{$^a$ This corresponds to model \eqref{eq: lme1} and ANOVA table \ref{tb: ANOVA1} in Web Appendix A.\\ $^b$ This corresponds to model \eqref{eq: lme2} and ANOVA table \ref{tb: ANOVA2} in Web Appendix A.\\
$^c$ This corresponds to model \eqref{eq: lme3} and ANOVA table \ref{tb: ANOVA2} in Web Appendix A.\\}
\end{table}

\clearpage

\section*{Web Appendix B: Clarifications on the correlated behavior-related latent variables between two raters}

In our framework, we assume the raters have two sets of binary actions during the decision process, (a) in Step I, whether or not to encounter uncertainty, (b) in Step II, whether or not to make a correct decision. Rather than modeling the binary outcomes directly, we introduce latent variables to facilitate appropriate modeling of potential between-rater correlations. For example, in Step I we denote $U_{ij}$ the Rater $i$'s binary status about ``uncertainty'' on rating subject $j$, $i=1,2$, $j=1,..., N$ ($U_{ij}=1$ if encountering uncertainty, otherwise, $U_{ij}=0$). We introduce continuous latent variable $L_{ij}^U$ corresponding to $U_{ij}$, with $U_{ij}=I(L^U_{ij}>0)$. 
It follows that 
\begin{align*}
    \begin{bmatrix}L_{1j}^U \\ L_{2j}^U\end{bmatrix} \sim BVN\left(\begin{bmatrix}\mu^U_{1} \\ \mu^U_{2} \end{bmatrix},\begin{bmatrix} 1 & \rho_U \\ \rho_U & 1\end{bmatrix}\right),
\end{align*}
where marginally $L_{ij}^U\sim N(\mu_i^U, 1)$, $\mu_i^U$ is the rater-specific effect, quantifying the capacity of feeling \textit{uncertain}; and $\rho_U$ is the correlation coefficient parameter between the two latent variables.  

Further, for rater $i$'s probability to encounter uncertainty, $p_i$, it is easy to show that
\begin{align*}
    p_i
    &=P(U_{ij}=1)=P(L_{ij}^U>0)=1-P(L_{ij}^U\le0)\\
    &=1-P\left(L_{ij}^U-\mu_i^U<-\mu_i^U\right)=1-\Phi\left(-\mu_i^U\right)=\Phi\left(\mu_i^U\right).
\end{align*}
Therefore, with $\mu_i^U=\Phi^{-1}(p_i)$, it will be consistent with our assumption that $U_{1j}$ and $U_{2j}$ are both Bernoulli distributed, with not necessarily the same probability parameters $p_1$ and $p_2$. By adopting latent variables with normally distributed errors, we can easily impose some correlation between $L_{1j}^U$ and $L_{2j}^U$, and hence, appropriately define the underlying correlation behind the binary outcomes $U_{1j}$ and $U_{2j}$. 

In Step II, we denote $C_{ij}$ the Rater $i$'s binary status about whether or not making a correct decision on rating subject $j$, $i=1,2$, $j=1,..., N$ ($C_{ij}=1$ if making a correct decision aligned with the underlying true outcome of the rating subject, otherwise, $C_{ij}=0$). Similarly, continuous latent variable $L_{ij}^C$ is introduced for $C_{ij}$, with $C_{ij}=I(L^C_{ij}>0)$. And it similarly follows that 
\begin{align*}
    \begin{bmatrix}L_{1j}^C \\ L_{2j}^C\end{bmatrix} \sim BVN\left(\begin{bmatrix}\mu^C_{1} \\ \mu^C_{2} \end{bmatrix},\begin{bmatrix}1 & \rho_C \\ \rho_C & 1\end{bmatrix}\right),
\end{align*}
where $L_{ij}^C\sim N(\mu_i^C, 1)$, $\mu_i^C$ and $\rho_C$ are the rater-specific means and the correlation coefficient for the latent variables in this Step II.  

Further, for the rater-specific misclassification rate parameter $m_i$, we can in the same way show that
\begin{align*}
    1-m_i
    &=P(C_{ij}=1)=P(L_{ij}^C>0)\\
    &=1-P\left(L_{ij}^C-\mu_i^C<-\mu_i^C\right)=\Phi\left(\mu_i^C\right).
\end{align*}
Hence, with $\mu_i^C=\Phi^{-1}(1-m_i)$, $C_{1j}$ and $C_{2j}$ are Bernoulli distributed with probability parameters $1-m_1$ and $1-m_2$. In this way, the underlying correlation behind the binary outcomes $C_{1j}$ and $C_{2j}$ is also properly defined. 

\clearpage

\section*{Web Appendix C: Data generation with correlated latent variables}
In this section, we clarify the mathematics in our simulation framework and the detailed data-generating mechanism. For $N$ rating subjects, we firstly generate the vector of their underlying truth $T$ with the prevalence parameter $\theta$. For subject $j$, $T_j=1$ with probability $\theta$ and $T_j=-1$ with probability $1-\theta$. We then define $(Y_{1}, Y_{2})$ as two vectors of rating outcomes given by Rater 1 and Rater 2, respectively (like the true membership $T_j$, $Y_{ij}\in \{-1,1\}$, for $i=1,2$ and $j=1,...N$, represent the negative or positive judgments in typical diagnostic questions). Define $(U_{1}, U_{2})$ to be Rater 1 and Rater 2's unobserved binary indicators about whether he/she feels uncertain about the question or not ($U_{ij}=1$ for uncertainty and $U_{ij}=0$ for certainty). The probability to feel uncertain about an item for rater $i$ is denoted as $p_i=P(U_i=1)$ for $i=1,2$. We then define continuous latent variables $L^U_{1}$ and $L^U_{2}$ to realize the correlated generation of $U_{1}$ and $U_{2}$. We simulate $(L^U_{1}, L^U_{2})$ from a bivariate normal distribution with mean $(\Phi^{-1}(p_1), \Phi^{-1}(p_2))$, unit variance, and covariance equal to $\rho_U$, where $\rho_U$ is a nonnegative correlation coefficient that also indirectly quantifies the correlation between $U_1$ and $U_2$. Note that $(U_1, U_2)=(I(L_1^U>0), I(L_2^U>0))$. If uncertainty is encountered, Rater $i$ will have probability $m_i$ to get the wrong guess or an opposite decision against the truth for $i=1,2$. We believe that the lower the $p_i$ and $m_i$ are, the more professional the rater $i$ will be. In the following steps, data generation will vary based on the number of raters that encounter uncertainty on the specific item. 

\subsection*{Scenario 1}
When both raters feel certain about item $j$, or $U_{1j}=U_{2j}=0$, perfect judgments can be made, so we generate $Y_{1j}=Y_{2j}=T_j$ for this item. 

\subsection*{Scenario 2}
When both raters feel uncertain about item $j$, or $U_{1j}=U_{2j}=1$, we define $(C_1, C_2)$ as Rater 1 and 2's indicators about whether they make a correct guess on this difficult question, where $C_{ij}=1$ for correct decisions and $C_{ij}=-1$ for wrong guesses. Recall the discussion of the correlated rater behaviors, we introduce another set of latent variables $(L_1^C, L_2^C)$ as the controllers of the correctness of the decision given a hard question has been encountered ($L_1^C>0$ for judgment consistent with the truth and $L_1^C<0$ for an opposite decision against truth). Then, we simulate $(L_1^C, L_2^C)$ from a bivariate normal distribution with mean $(\mu_1=\Phi^{-1}(1-m_1), \mu_2=\Phi^{-1}(1-m_2))$, unit variance, and covariance matrix equal to $\rho_C$, where $1-m_1=P(L_1^C>0)=P(C_1=1)$ and $1-m_2=P(L_2^C>0)=P(C_2=1)$ are the respective marginal probability to make the correct guess for each rater, and $\rho_C$ is some correlation measure of raters' ability to make right guesses when coping with difficult judgments. Finally, we generate $(Y_{1j}, Y_{2j})$ based on $(C_{1j}T_{j}, C_{2j}T_{j})$ for $j=1,...,N$, where $C_{ij}=1$ when $L_{ij}^C>0$ and $C_{ij}=-1$ when $L_{ij}^C<0$.

\subsection*{Scenario 3}
When one rater feels uncertain and the other rater feels certain about question $j$, without loss of generality, we assume $U_{1j}=0$ and $U_{2j}=1$. This allows us to firstly generate $Y_{1j}=T_j$. $U_{1j}=0$ implies $C_{1j}=1$ and $L^C_{1j}>0$, so, we generate $L^C_{1j}$ from truncated normal distribution $N(\mu_1, 1;L^C_{1j}>0)$. Then, simulate $L_{2j}^C$ from the conditional distribution $L^C_{2j}|L^C_{1j}=l_1 \sim N(\mu_2+\rho_C l_1-\rho_C \mu_1,~ 1-\rho_C^2)$, with $\mu_1$ and $\mu_2$ defined above (The corresponding derivation can be found in Appendix). Finally, we generate $C_{2j}$ based on $L_{2j}^C$'s sign and get $Y_{2j}=C_{2j}T_{j}$. In symmetry, for the case of $U_{1j}=1$ and $U_{2j}=0$, we will simulate $Y_{2j}=T_j$ and $Y_{1j}=C_{1j}T_{j}$, where $C_{1j}$ is decided by the sign of $L_{1j}^C$, and $L_{1j}^C$ is sampled from $L^C_{1j}|L^C_{2j}=l_2 \sim N(\mu_1+\rho_C l_2-\rho_C \mu_2,~ 1-\rho_C^2)$.

After generating the decision outcomes for the two raters with respect to each rating subject, we can tabulate this $(Y_1, Y_2)$ into the $2\times 2$ contingency table to calculate each IRA statistic.

\section*{Web Appendix D: Derivations of the $4\times 4$ ``true'' probability table}

Below, we show the derivations for the cell components of the $4\times 4$ ``truth'' table. Denote the probability of both raters encountering uncertainty as $U_{11}$, the probability of Rater 1 feeling uncertain but Rater 2 feeling certain as $U_{10}$, the probability of Rater 1 feeling certain but Rater 2 feeling uncertain as $U_{01}$, and the probability of both encountering certainty as $U_{00}$. Based on the parameters we defined in our data-generating framework, these four probabilities can be calculated as follows. Note that $U_{10}$ is not necessarily the same as $U_{01}$ due to the possibly distinct rater characteristics. 
\begin{align*}
    &U_{11}=P(U_1=1, U_2=1)=P(L^U_1>0, L^U_2>0)=\int^\infty_0\int^\infty_0f_{L^U_1,L^U_2} dl^U_1dl^U_2,\\
    &U_{10}=P(U_1=1, U_2=0)=P(L^U_1>0, L^U_2<0)=\int^0_{-\infty}\int^\infty_0f_{L^U_1,L^U_2} dl^U_1dl^U_2,\\
    &U_{01}=P(U_1=0, U_2=1)=P(L^U_1<0, L^U_2>0)=\int^\infty_0\int^0_{-\infty}f_{L^U_1,L^U_2} dl^U_1dl^U_2,\\
    &U_{00}=1-U_{11}-U_{10}-U_{01}.
\end{align*}
where $U_1$ and $U_2$ are the binary indicators of encountering uncertainty or not for Rater 1 and 2, and $f_{L^U_1,L^U_2}$ is the density function of the bivariate normal distribution regarding the continuous latent variables $(L^U_1,L^U_2)$ as defined in Web Appendix C. Also, note that $U_{11}$, $U_{10}$, $U_{01}$, and $U_{00}$ are all functions of $p_1$, $p_2$, and $\rho_U$.

For the case that both raters feel uncertain about the subject (the upper-left $2\times 2$ block in the true table), we define the four joint probabilities about the two raters' correctness on the decision-making as $C_{11}$, $C_{10}$, $C_{01}$, and $C_{00}$ (with similar rules of subscript definitions). Similarly, we can calculate them through
\begin{align*}
    &C_{11}=P(C_1=1,C_2=1)=P(L^C_1>0, L^C_2>0)=\int^\infty_0\int^\infty_0f_{L^C_1,L^C_2} dl^C_1dl^C_2,\\
    &C_{10}=P(C_1=1,C_2=-1)=P(L^C_1>0, L^C_2<0)=\int^0_{-\infty}\int^\infty_0f_{L^C_1,L^C_2} dl^C_1dl^C_2,\\
    &C_{01}=P(C_1=-1,C_2=1)=P(L^C_1<0, L^C_2>0)=\int^\infty_0\int^0_{-\infty}f_{L^C_1,L^C_2} dl^C_1dl^C_2,\\
    &C_{00}=1-C_{11}-C_{10}-C_{01},
\end{align*}
where $C_1$ and $C_2$ are the binary indicators of making a correct decision or not for Rater 1 and 2, and $f_{L^C_1,L^C_2}$ is the joint density for bivariate normal-distributed $(L^C_1,L^C_2)$ as defined in Web Appendix C. Also, note that $C_{11}$, $C_{10}$, $C_{01}$, and $C_{00}$ are all functions of $m_1$, $m_2$, and $\rho_C$.

Denote Rater 1 and 2's respective decision as $Y_1, Y_2\in \{1, -1\}$ (``positive" or ``negative"). Recall that $\theta$ is the probability of a ``positive" outcome. Then, we have
\begin{align*}
    &P(Y_1=1, Y_2=1 | U_1=1, U_2=1)=\theta C_{11}+(1-\theta) C_{00},\\
    &P(Y_1=1, Y_2=-1 | U_1=1, U_2=1)=\theta C_{10}+(1-\theta) C_{01},\\
    &P(Y_1=-1, Y_2=1 | U_1=1, U_2=1)=\theta C_{01}+(1-\theta) C_{10},\\
    &P(Y_1=-1, Y_2=-1 | U_1=1, U_2=1)=\theta C_{00}+(1-\theta) C_{11}.
\end{align*}

For the case of the lower-left $2\times 2$ block ($U_1=0, U_2=1$, $C_1=1$ and $L^C_1>0$), we define a conditional quantity such that 
\begin{align*}
    C_{2|1}=P(C_2=1|C_1=1)=\frac{P(C_1=1, C_2=1)}{P(C_1=1)}=\frac{P(L^C_1>0,L^C_2>0)}{P(L_1^C>0)}=\frac{\int^\infty_0\int^\infty_0f_{L^C_1,L^C_2} dl^C_2dl^C_1}{\int^\infty_0f_{L^C_1} dl^C_1}
\end{align*}
where $f_{L^C_1}$ is the marginal density for $L_1^C$, as defined in Web Appendix C. Then, we have
\begin{align*}
    &P(Y_1=1, Y_2=1 | U_1=0, U_2=1)=\theta C_{2|1},\\
    &P(Y_1=1, Y_2=-1 | U_1=0, U_2=1)=\theta(1-C_{2|1}),\\
    &P(Y_1=-1, Y_2=1 | U_1=0, U_2=1)=(1-\theta)(1-C_{2|1}),\\
    &P(Y_1=-1, Y_2=-1 | U_1=0, U_2=1)=(1-\theta)C_{2|1}.
\end{align*}

Similarly, for the case in the upper-right block, we symmetrically define $C_{1|2}$ and write that 
\begin{align*}
    C_{1|2}=P(C_1=1|C_2=1)=\frac{P(C_1=1, C_2=1)}{P(C_2=1)}=\frac{P(L^C_1>0,L^C_2>0)}{P(L^C_2>0)}=\frac{\int^\infty_0\int^\infty_0f_{L^C_1,L^C_2} dl^C_1dl^C_2}{\int^\infty_0f_{L^C_2} dl^C_2}
\end{align*}
where $f_{L^C_2}$ is the marginal density for $L_1^C$.
Then, we have
\begin{align*}
    &P(Y_1=1, Y_2=1 | U_1=1, U_2=0)=\theta C_{1|2},\\
    &P(Y_1=1, Y_2=-1| U_1=1, U_2=0)=(1-\theta)(1-C_{1|2}),\\
    &P(Y_1=-1, Y_2=1| U_1=1, U_2=0)=\theta(1-C_{1|2}),\\
    &P(Y_1=-1, Y_2=-1| U_1=1, U_2=0)=(1-\theta)C_{1|2},\\
\end{align*}

Finally, it is easy to insert these probabilities of response combinations conditional on the ``uncertainty" memberships into the $4\times 4$ ``truth" table.

\clearpage

\section*{Web Appendix E: Additional simulation results under the assumption of no certainty/uncertainty classification}

The simulation framework and the ``true'' IRA definition introduced in the main paper are based on the assumption of the existence of certainty and uncertainty (i.e., the Step-I decision in the rating process) as well as the fact that certainty means absolutely correct decision. The classification of certainty and uncertainty was never considered in the development of IRA methods except for Gwet's $\text{AC}_1$. This raised some suspicions from the reviewers' about the fairness of our simulation design. However, our proposed generalized simulation framework has the flexibility to reduce to a framework without considering the Step-I decision. Specifically, when $p_1=p_2=1$, the two raters always feel \textit{uncertain}, and the rating process only involve the Step-II decision about making correct decision or not. In this case, with the ``truth'' table reduced to only the upper-left $2\times 2$ block, the ``true'' chance-corrected IRA can be expressed as,
\begin{align*}
    K=\frac{\gamma^2 (C_{11}+C_{00})}{1-(1-\gamma^2)(C_{11}+C_{00})},
\end{align*}
where $C_{11}+C_{00}=p_a$ is the total agreement proportion, and $\gamma^2$ is the proportion of true agreement within the total agreement. When doing the simulations, we also summarized the comparisons when $p_1=p_2=1$ as some supplementary results. Under this case of only having the Step-II decision, the $\rho_U$ parameter will no longer appear, and if varying other parameters based on the simulation settings introduced in the main paper, there are totally $9\times 5\times 5\times 5\times 4=4,500$ simulation settings. We parallelly show the simulation results based on the presenting order of the main simulation results.

\subsection*{Overall evaluations across all settings}

\begin{figure}[htbp!]
\centering
\includegraphics[width=1.1\textwidth]{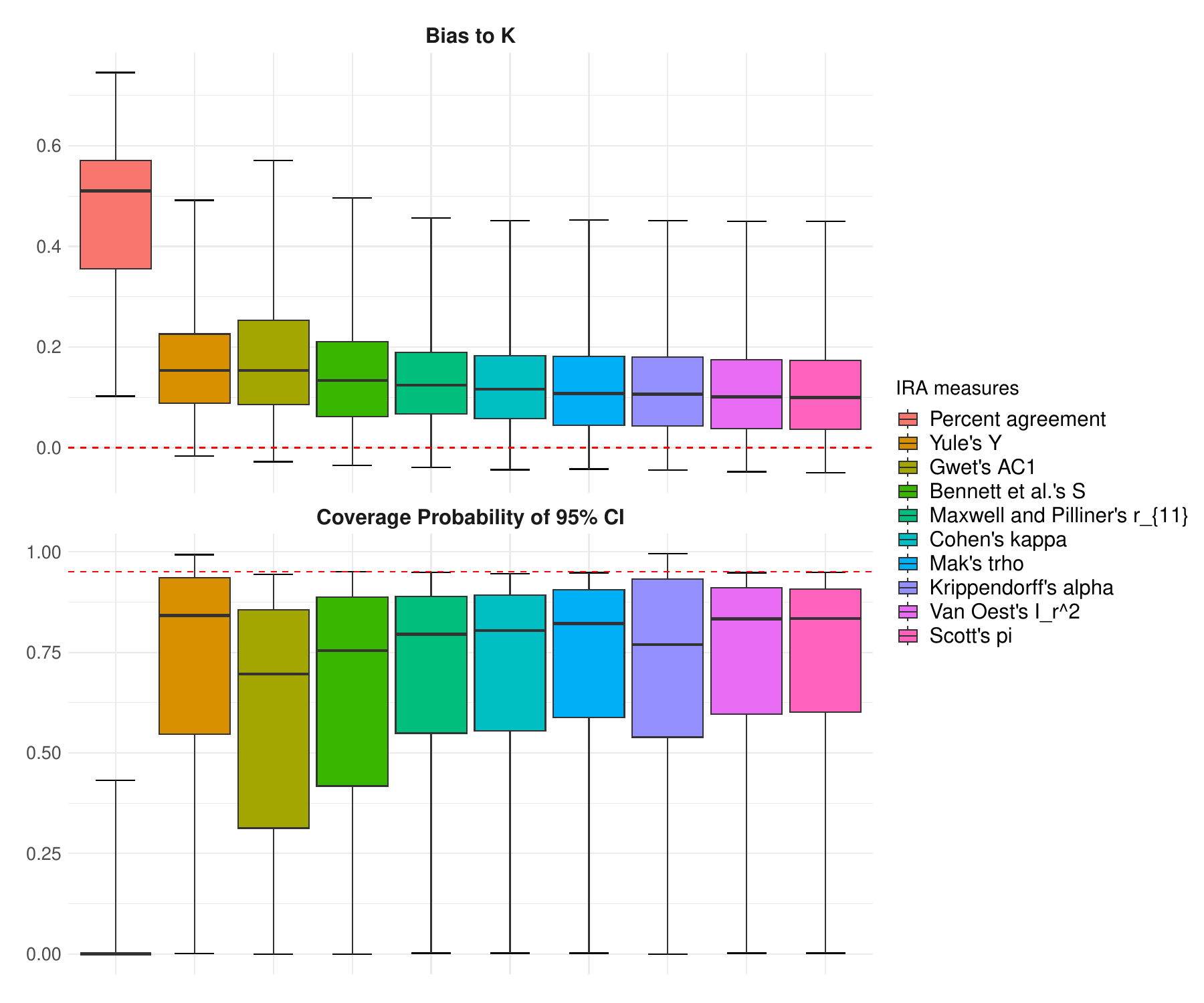}
\caption{Across-scenario overall bias towards the ``true'' interrater agreement $K$ and overall coverage probability using $95\%$ confidence intervals for $10$ interrater agreement measures over the $4,500$ simulation settings by assuming $p_1=p_2=1$. From the bottom to the top, the five summary statistics in the box plots are the 2.5 percentile, the first quartile, the median, the third quartile, and the 97.5 percentile.} \label{fig: overall_sp} 
\end{figure}

Figure \ref{fig: overall_sp} shows the biases and coverage probabilities of different IRA measures compared to the ``true'' chance-corrected IRA $K$ among all the setting assuming no certainty/uncertainty classification (i.e., $p_1=p_2=1$). Akin to the main paper, the $10$ IRA measures were ranked based on the magnitudes of median bias values from left to right, and we kept this order through all the result presentations. Over the full parameter constellations under this reduced data-generating framework, Scott's $\pi$ and those similarly-performed method cluster in the main paper show relatively smaller median bias. However, all the IRA methods overall show positive bias towards the benchmark $K$. This indicates that the benchmark quantity $K$ generally corrects down more ``chance agreement'' than the existing methods. 

\begin{figure}[htbp!]
\centering
\includegraphics[width=1\textwidth]{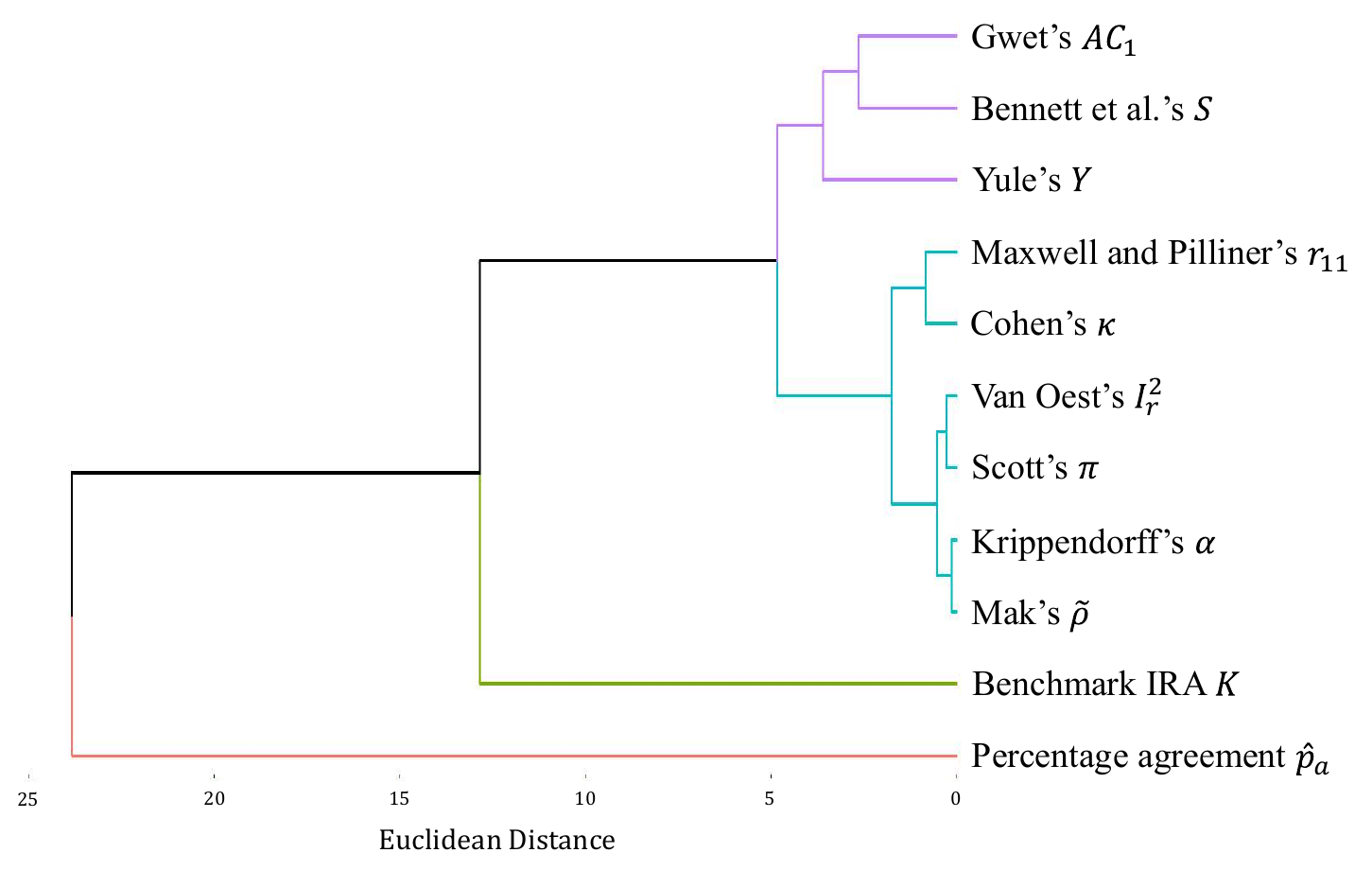}
\caption{Agglomerative hierarchical clustering of the $10$ IRA measures and the ``true'' interrater agreement $K$ based on $4,500$ simulation settings assuming $p_1=p_2=1$.} \label{fig: clustering_sp} 
\end{figure}

Figure \ref{fig: clustering_sp} assesses the closeness among the $10$ interrater agreement methods as well as the benchmark measure $K$ based on their estimates over all the simulation settings given $p_1=p_2=1$. The same as the clustering analysis results in the main paper, the hierarchical clustering indicates that the percent agreement $\hat{p}_a$ itself forms a cluster, $Y$, $S$, and $\text{AC}_1$ are likely to be in the same cluster, and $r_{11}$, $\kappa$, $\alpha$, $\tilde{\rho}$, $\pi$, and $I_r^2$ form a cluster. Differetly, $K$ at this time is not close to either IRA method in this overall analysis.

\subsection*{Bias evaluations at fixed levels of key factors}

\begin{figure}[htbp!]
\centering
\includegraphics[width=1\textwidth]{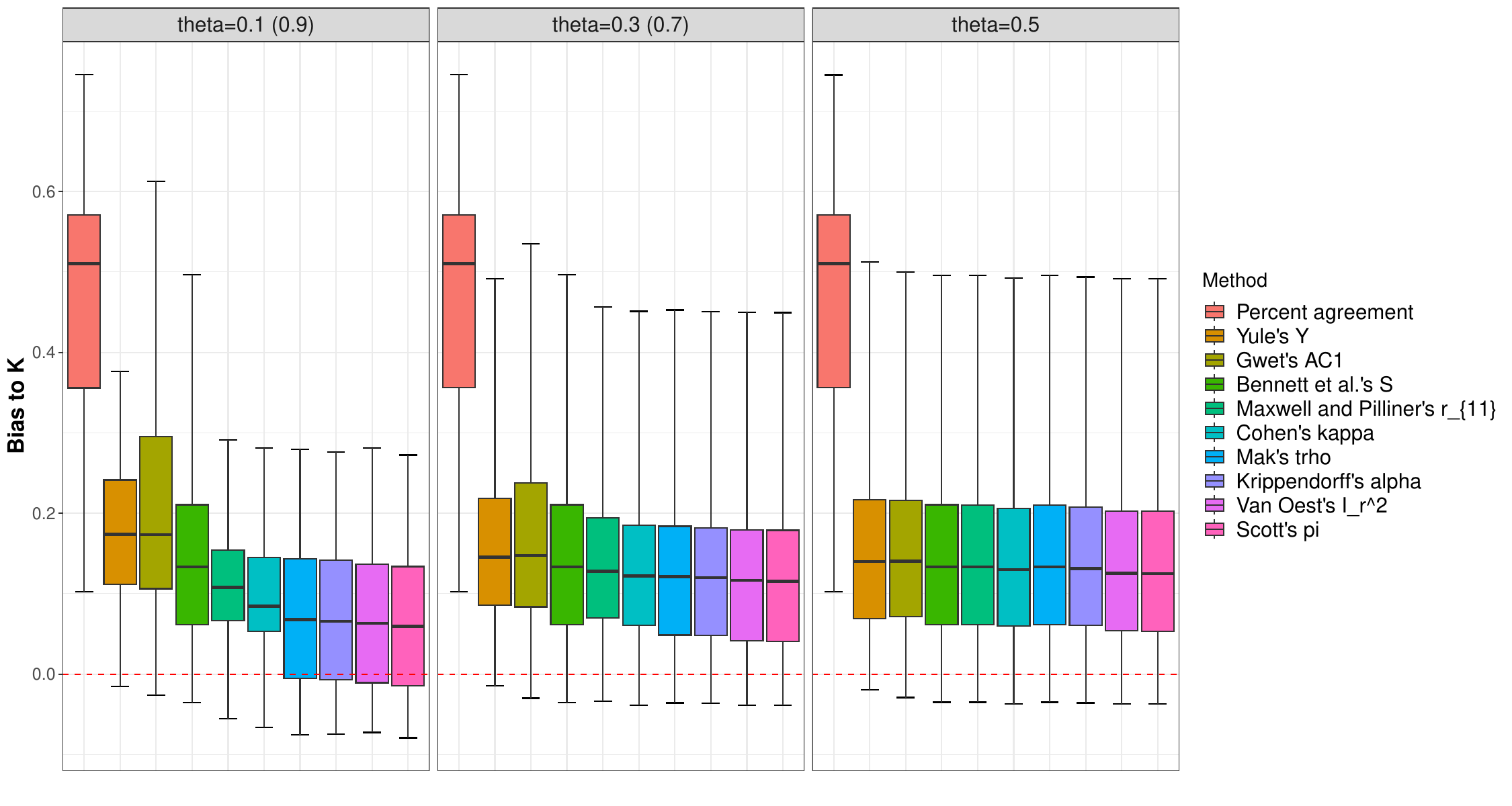}
\caption{Across-scenario overall bias to ``true'' interrater agreement $K$ for $10$ interrater agreement measures at different outcome prevalence levels.} \label{fig: theta_bias_sp} 
\end{figure}

In this part, we parallel the exploration of the bias performance of different IRA methods when fixing one or more of the key simulation parameters at specific levels. Figure \ref{fig: theta_bias_sp} shows the overall comparison at extreme to non-extreme prevalence levels. Across different panels, it seems that all of the IRA methods uniformly overestimate $K$ no matter how extreme the outcome prevalence is. This shows an apparent distinction compared to the impactful role of $\theta$ shown in the main results.

\begin{figure}[htbp]
\centering
\includegraphics[width=1\textwidth]{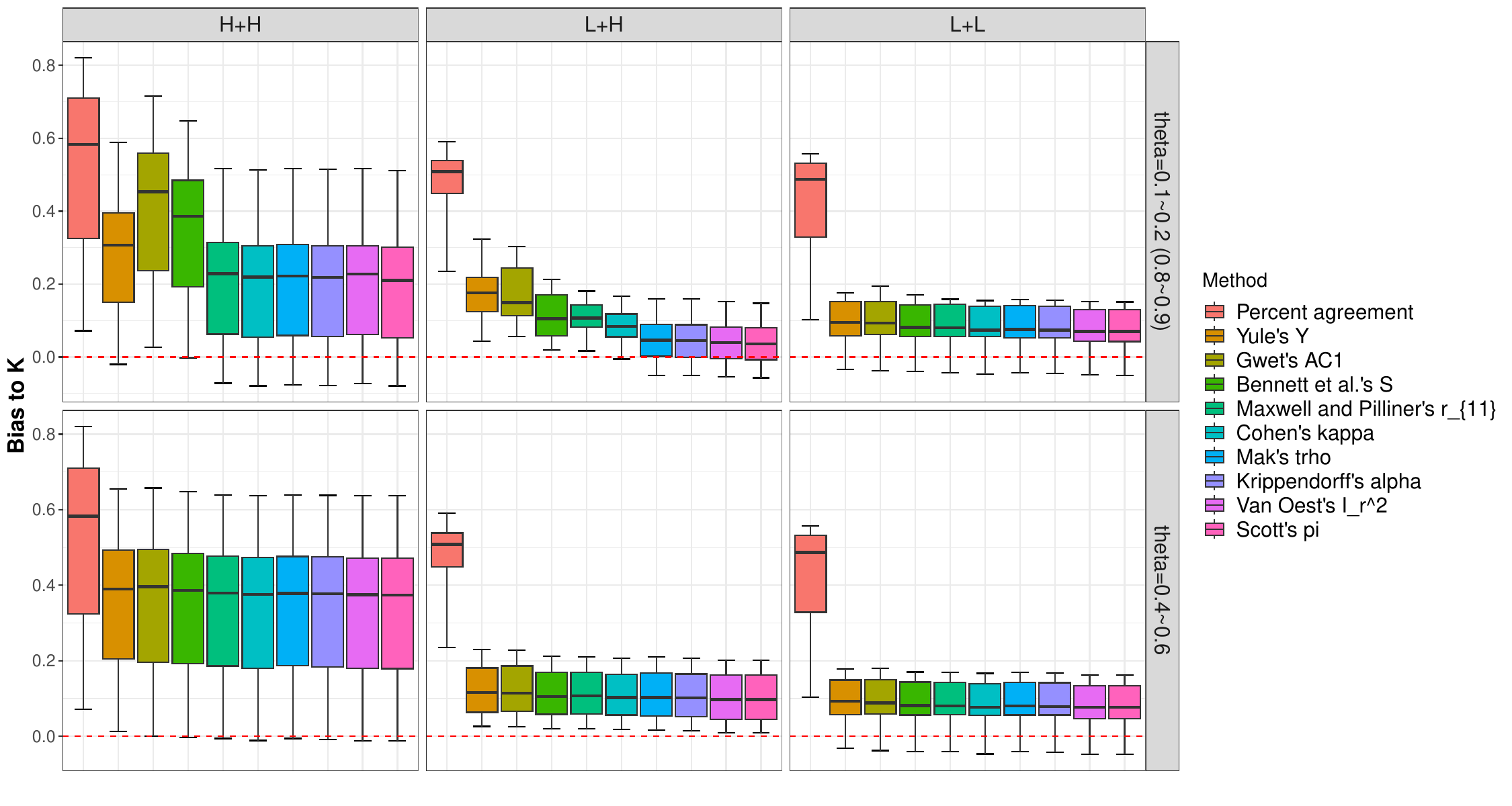}
\caption{Across-scenario overall bias to ``true'' interrater agreement $K$ for $10$ interrater agreement measures at different joint rater-professionalism levels and typical outcome prevalence levels.} \label{fig: JPL_bias_sp} 
\end{figure}

Although $\theta$ is not as influential as in the general certainty/uncertainty cases in the main paper, we still show the results in a similar way as in the main simulation. Figure \ref{fig: JPL_bias_sp} shows the overall comparison by fixing rater misclassification rates at certain hypothetical levels. We assume that those \textit{professional} raters tend to have $m_i$ low, $i=1,2$. We similarly define two levels of rater expertise within our settings: raters with ``high" professionalism ($m_i=0.1~\text{and}~0.2$, denoted as ``H'') and raters with ``low" professionalism ($m_i=0.4~\text{and}~0.5$, denoted as ``L''). We present three levels of rater-professionalisms in Figure \ref{fig: JPL_bias_sp}, from two ``high''-profession raters (H+H) to two ``low''-profession raters (L+L). With the joint professional levels decrease, all the methods have their bias to $K$ decreased.

\begin{figure}[htbp]
\centering
\includegraphics[width=1\textwidth]{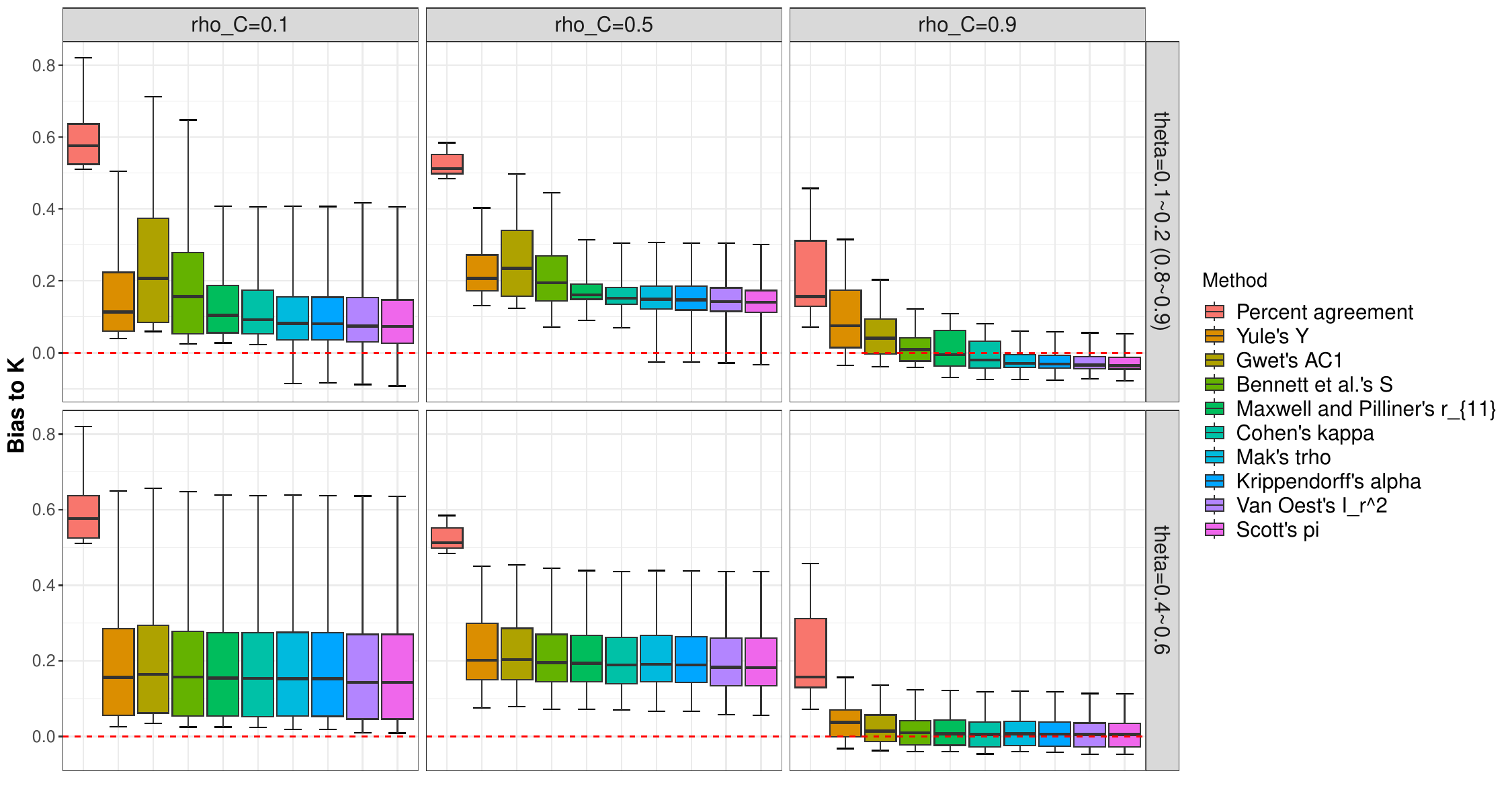}
\caption{Across-scenario overall bias to ``true'' interrater agreement $K$ for $10$ interrater agreement measures at different aggregate rater correlation levels and typical outcome prevalence levels.} \label{fig: rho_bias_sp} 
\end{figure}

Finally, we vary $\rho_C$ in Figure \ref{fig: rho_bias_sp} to approximate low to high levels of overall rater behavioral correlation. Surprisingly, when $\rho_C$ is close to 1, then that cluster of similarly-performed methods (i.e., from Maxwell and Pilliner's $r_{11}$ to Scott's $\pi$) show minimal bias towards $K$ across the settings varied by the other parameters. This phenomena can be observed for both extreme and non-extreme outcome prevalence.

\end{document}